\documentclass{article}
\newif\ifarxiv
\ifdefined\arxivversion\arxivtrue\fi
\arxivtrue
\ifdefined\pdfsuppressptexinfo\pdfsuppressptexinfo=-1\fi
\usepackage{iclr2027_conference,times}

\usepackage{amsmath,amsfonts,bm}

\def\eqref#1{equation~\ref{#1}}

\def\1{\bm{1}}

\DeclareMathAlphabet{\mathsfit}{\encodingdefault}{\sfdefault}{m}{sl}
\SetMathAlphabet{\mathsfit}{bold}{\encodingdefault}{\sfdefault}{bx}{n}

\usepackage{url}

\usepackage{graphicx}
\usepackage{booktabs}
\usepackage{amsmath, amssymb}
\usepackage{algorithm}
\usepackage{algorithmic}
\usepackage{bbm}
\usepackage{diagbox}
\usepackage{wrapfig}
\usepackage{multirow}
\usepackage[table]{xcolor}
\usepackage{hyperref}

\definecolor{tbgreen}{HTML}{C7EABB}
\definecolor{tbgrey}{HTML}{E8E8E8}

\definecolor{citationblue}{HTML}{245A81}
\definecolor{appendixteal}{HTML}{27686A}
\hypersetup{
    colorlinks=true,
    citecolor=citationblue,
    linkcolor=black,
    urlcolor=citationblue
}

\newcommand\minisection[1]{\vspace{2mm}\noindent \textbf{#1}}

\renewcommand{\floatpagefraction}{0.85}

\title{ReCoVR: Closing the Loop in Interactive \\ Composed Video Retrieval}
\ifarxiv
\iclrfinalcopy
\author{Bingqing Zhang, Yi Zhang, Zhuo Cao, Yang Li, Xue Li, Jiajun Liu, Sen Wang \\
The University of Queensland, CSIRO Data61, Australia \\
\texttt{bingqing.zhang@uq.edu.au}}
\else
\author{Anonymous authors}
\fi

\begin{document}
\maketitle
\ifarxiv\lhead{Preprint}\fi

\begin{abstract}
Composed video retrieval (CoVR) retrieves a target video from a reference video and a modification text, yet a single query often leaves the target underspecified and existing methods offer no way to refine it. We formalize \textit{interactive composed video retrieval} (ICoVR), a multi-turn extension of CoVR in which users refine their intent through natural-language feedback. Adapting interactive text-to-video retrieval methods to ICoVR exposes two structural weaknesses: they route all feedback through a \textit{single retrieval channel}, and they operate \textit{open-loop}, updating the query from feedback without checking whether their own rankings are drifting or stagnating. After a turn that fails to improve the target rank, the strongest adapted baseline improves on the next turn in only 30.12\% of cases. We propose \textbf{ReCoVR}, a training-free framework that closes this loop by treating the system's own ranking trajectory as evidence alongside user feedback. An Intent Pathway routes each feedback turn to text-to-video and composed retrieval channels, and a Reflection Pathway accumulates constraints from feedback and uses ranking history to demote rejected or stagnant results. On WebVid-CoVR-Test, ReCoVR reaches 74.30\% Recall at 1 (R@1) after one feedback turn and 90.22\% after five, and its margin over the strongest baseline widens from 7.2 to 15.2 points. The gains carry over to Dense-WebVid-CoVR and FineCVR, and evaluation with real human feedback shows the same improvement trend.
\end{abstract}

\section{Introduction}
\label{sec:intro}

\begin{figure}[tbp]
    \centering
    \includegraphics[width=1.0\linewidth]{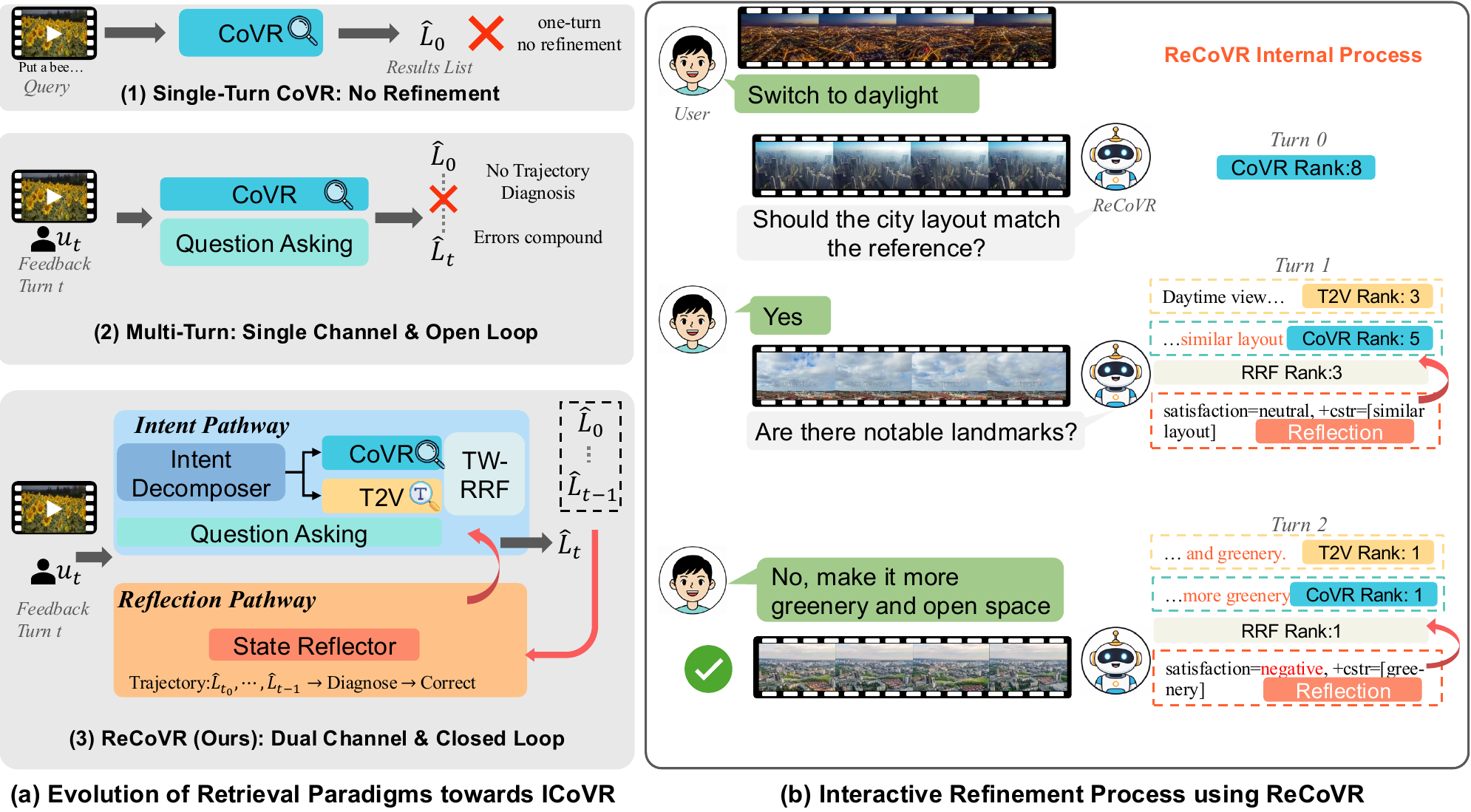}
    \vspace{-2pt}
    \caption{\textbf{Overview of Interactive Composed Video Retrieval (ICoVR) and ReCoVR.} \textbf{(a) Paradigm Comparison:} (1) Traditional CoVR is single-turn only. (2) Naive multi-turn methods use a single channel and an \textit{open loop}: they update from user feedback without diagnosing past ranked results, allowing errors to compound. (3) Our framework \textbf{ReCoVR} uses an Intent Pathway that routes queries to composed video retrieval (CoVR) and text-to-video (T2V) channels, and a closed-loop Reflection Pathway that diagnoses past retrieval trajectories. \textbf{(b) Interaction Example:} Across turns, ReCoVR extracts constraints from feedback and ranks videos to reach the target.}
    \label{fig:intro}
    \vspace{-3pt}
\end{figure}

Composed video retrieval (CoVR)~\citep{webvid_covr, webvid_covr2} searches for a target video using a reference video and a modification text, with steady progress in datasets~\citep{webvid_covr2, densewebvidcovr, finecvr} and methods~\citep{covr_enrich, finecvr}. Existing approaches, however, assume a single query-response cycle. This assumption is restrictive for CoVR because a reference video and a short modification text often admit multiple plausible targets. Consider a user with a reference video of a city at night who asks the system to ``\texttt{switch to daylight}.'' The system returns a daytime scene, but of a different city with dense skyscrapers instead of the low-rise layout in the reference. The user refines the query with ``\texttt{keep a similar layout},'' and the system now returns a matching layout at night, discarding the daylight constraint it had already satisfied. Under the single-turn paradigm (Fig.~\ref{fig:intro}(a.1)), each refinement is a new query treated in isolation, so earlier progress can be undone without notice. In realistic visual search, users supply such missing constraints progressively after observing imperfect results~\citep{marchionini2006exploratory}, which motivates extending CoVR to multi-turn interaction.

Interactive retrieval has been explored in neighboring domains. In interactive text-to-video retrieval (I-TVR), prior methods typically ask clarifying questions, collect user answers, and use them to refine the query~\citep{ivr_auto,merlin,umivr}. Related ideas also appear in visual dialog~\citep{guo2018dialog} and interactive composed image retrieval~\citep{fashionntm, mai, luo2025imagescope}. However, these settings do not address the composed video case, where the system must jointly maintain a visual reference and an evolving modification intent across turns. We therefore formulate \textbf{Interactive Composed Video Retrieval (ICoVR)}. The user starts with a standard CoVR query at Turn~0 and can refine the target through natural-language feedback from Turn~1 onward (Fig.~\ref{fig:intro}(b)). This feedback can be spontaneous or answer an optional clarifying question from the system. To evaluate this setting, we adopt a controlled simulated-user protocol and adapt representative I-TVR methods as baselines.

Once transferred to ICoVR, these baselines reveal two structural weaknesses (Fig.~\ref{fig:intro}(a.2)). First, they rely on a \textit{single retrieval channel}. User feedback either describes a change to the current result or describes the target directly, yet a composed channel ties both forms to the reference video, even when a target description could be used directly by a text-to-video retriever. Second, they operate as \textit{open-loop} systems: their query updates are driven by user feedback, but they do not check whether their own retrieval trajectory is drifting, stagnating, or repeatedly returning rejected candidates. This failure is measurable. The strongest adapted baseline fails to improve the target rank in 54.25\% of its active feedback turns, and after such a turn it improves on the next executed turn in only 30.12\% of cases (Sec.~\ref{sec:main_results}). Feedback alone thus often leaves earlier retrieval failures unresolved. Our key insight is that \textit{the system's own ranking trajectory is an observable signal of these failures}: a top result that the user has rejected, or one that persists across turns, can be detected and corrected without any access to the target, using only rankings the system has already produced.

Building on this insight, we propose \textbf{ReCoVR} (\textbf{Re}flexive \textbf{Co}mposed \textbf{V}ideo \textbf{R}etrieval), a training-free closed-loop framework for ICoVR that reads its own retrieval trajectory alongside user feedback (Fig.~\ref{fig:intro}(a.3)). An \textbf{Intent Pathway} addresses the single-channel limitation by decomposing each feedback turn into a direct target description and a relative edit, routing them to complementary text-to-video (T2V) and CoVR channels. A \textbf{Reflection Pathway} updates semantic constraints from user feedback and uses ranking history to detect repeated results. It then lowers the ranks of rejected or stagnant candidates before fusion.
These pathways are supported by a dual memory bank that tracks semantic constraints and ranking history, and are unified by a time-weighted reciprocal rank fusion strategy that aggregates multi-turn evidence while emphasizing recent feedback. Because it requires no parameter updates, ReCoVR is backbone-agnostic and plugs into existing retrieval models.
In summary, our contributions are three-fold:
\begin{enumerate}

    \item We formulate \textbf{Interactive Composed Video Retrieval (ICoVR)}, a multi-turn extension of CoVR for progressive intent refinement through natural-language feedback, and identify single-channel execution and open-loop updating as weaknesses of adapted retrievers.

    \item We propose \textbf{ReCoVR}, a training-free framework that routes feedback to complementary retrieval channels and uses the ranking trajectory to correct both the next query, via accumulated constraints, and the current ranking, via rank-level demotion.

    \item ReCoVR outperforms adapted interactive baselines on three CoVR benchmarks, with an R@1 margin on WebVid-CoVR-Test that widens from 7.2 to 15.2 points over five turns. After a turn that fails to improve the target rank, it improves the rank on the next turn in 88.15\% of cases versus 30.12\% for the strongest baseline, and matched ablations attribute the gains to dual-channel routing, semantic reflection, and rank correction.
\end{enumerate}

\section{Related Work}
\label{sec:related_work}

\textbf{Composed Video Retrieval.}
Composed Video Retrieval (CoVR) extends Composed Image Retrieval~\citep{han2017automatic, vo2019composing, chen2020learning, cirr, fashioniq} to the video domain, where a target video is retrieved from a reference video and a modification text. Existing methods have advanced through large-scale caption-derived training data~\citep{webvid_covr, webvid_covr2}, visual-bias mitigation by separating preserved content from text-driven changes~\citep{covr_enrich, hud, cadp, refine}, and fine-grained temporal benchmarks or dense descriptions~\citep{egocvr, tfcovr, finecvr, densewebvidcovr, omnicvr}. The reason-aware CoVR-R~\citep{thawakar2026covr} predicts edit after-effects such as state transitions, action phases, and camera changes, but still retrieves from a single fixed reference--edit pair. ICoVR additionally requires correcting drift, stagnation, and repeated feedback violations across turns.

\textbf{Interactive Visual Retrieval.}
Interactive retrieval refines search through iterative user feedback. In the image domain, relevance feedback~\citep{rui1998relevance, wu2004willhunter} evolved into state aggregation and sequence modeling~\citep{guo2018dialog, cai2021ask, yuan2021conversational,fashionntm,wei2023conversational} and, recently, multi-turn reasoning with large multimodal models~\citep{chatir, plugir, mai,feng2025interactive, lu2025llava, bai2025chat, he2025chatting, qin2025human}. For videos, interactive text-to-video retrieval (I-TVR)~\citep{han2025ivcr} refines queries by appending answers~\citep{ivr_auto}, interpolating embeddings~\citep{merlin}, rewriting dialogue history~\citep{wu2024multimodal} or quantifying uncertainties~\citep{umivr}.
None of these methods addresses the composed video setting, which must jointly maintain a visual reference and an evolving modification text. Beyond interactive retrieval, negative relevance feedback demotes documents a user has rejected~\citep{wang2008study, bi2019conversational}, and self-reflective LLM agents revise later actions from earlier outcomes~\citep{yao2022react, shinn2023reflexion}. ReCoVR applies both ideas to interactive retrieval but reflects on a specific signal, the ranked-list trajectory of its own retrieval channels, and acts only through query constraints and rank-level corrections. Adapted I-TVR methods consume feedback but do not diagnose their past rankings.

\section{Method}
\label{sec:method}

\subsection{Problem Setup and ReCoVR Overview}
\label{sec:setup}

\begin{figure}[tbp]
    \centering
    \includegraphics[width=0.98\linewidth]{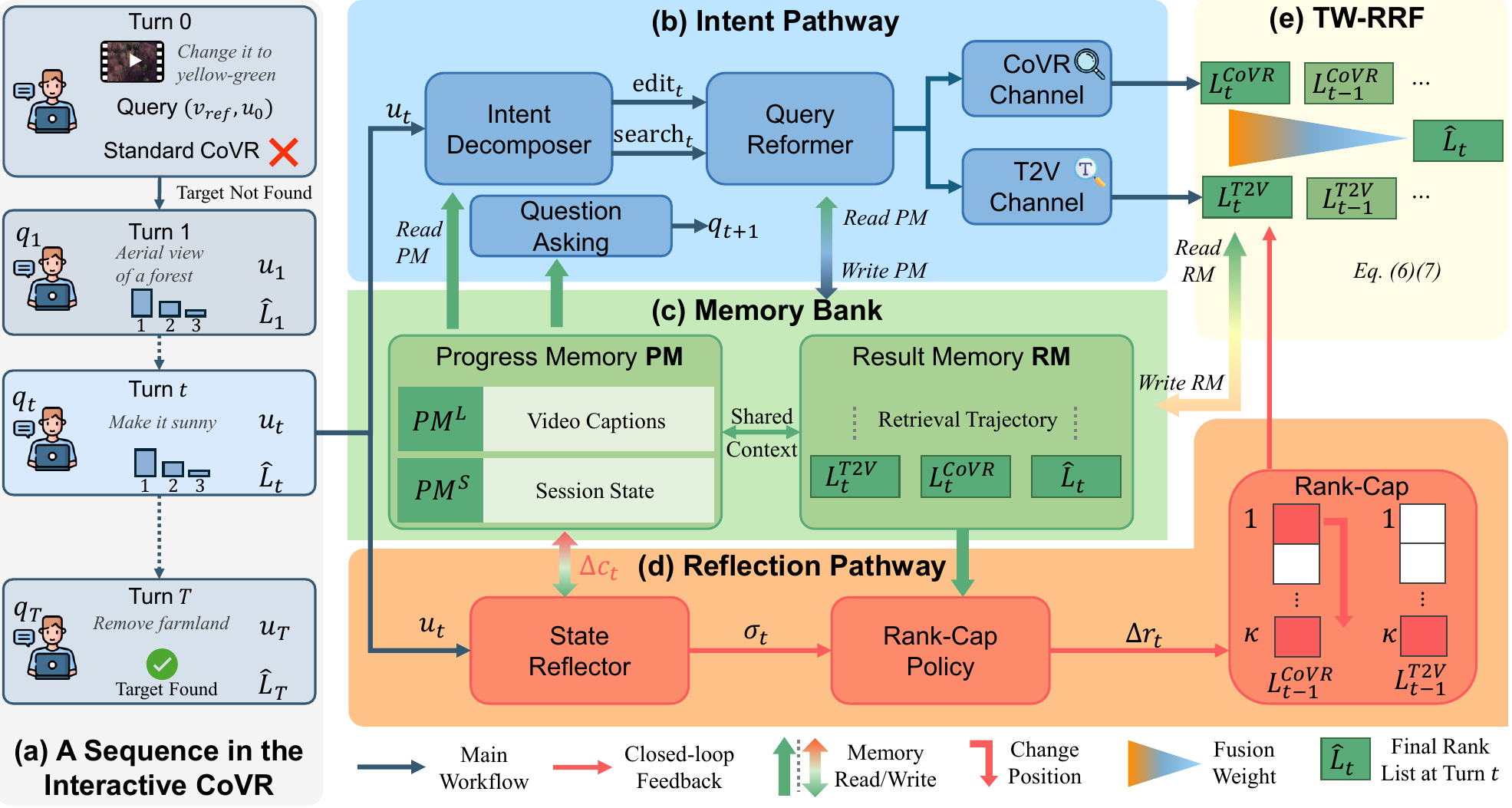}
    \vspace{-2pt}
    \caption{\textbf{Pipeline of ReCoVR.} When standard CoVR fails at Turn~0 \textbf{(a)}, the system enters an interactive loop. At each turn, the Intent Pathway \textbf{(b)} decomposes user feedback $u_t$ into dual-channel queries and generates a clarifying question $q_{t+1}$, while the Reflection Pathway \textbf{(d)} infers satisfaction $\sigma_t$ and constraints $\Delta c_t$ from the session state, then consults the retrieval trajectory to apply rank caps $\Delta r_t$. Both pathways read from and write to a shared Memory Bank \textbf{(c)} that tracks semantic state (PM) and ranked results (RM). Finally, TW-RRF \textbf{(e)} fuses candidates over a recency-weighted sliding window into the final ranked list $\hat{L}_t$.}
    \label{fig:method}
    \vspace{-3pt}
\end{figure}

\textbf{Interactive Composed Video Retrieval (ICoVR)} extends single-turn composed video retrieval~\citep{webvid_covr, webvid_covr2} to multiple turns (Fig.~\ref{fig:method}(a)). Given a reference video $v_{\mathrm{ref}}$, an initial modification text $u_0$, and a gallery $\mathcal{G}$, the user first issues a standard CoVR query at Turn~0 and receives an initial ranked list $\hat{L}_0$. When the target is not found, the user can continue the search from Turn~1 onward by providing natural-language feedback:
\begin{equation}
  \bigl\{(q_t,\;u_t,\;\hat{L}_t)\bigr\}_{t=1}^{T},
  \label{eq:icovr}
\end{equation}
where $T$ is the maximum number of feedback turns, $q_t$ is an optional system-generated clarifying question, $u_t$ is the user's feedback or refinement instruction, and $\hat{L}_t$ is the updated ranked list. The goal is to progressively align the ranked list with the user's intent over turns. The system never observes the target. When producing $\hat{L}_t$, its only information is the interaction history
\begin{equation*}
  \mathcal{H}_t = \bigl(v_{\mathrm{ref}},\; u_0,\; \{(q_i, u_i)\}_{i=1}^{t},\; \{\hat{L}_i\}_{i=0}^{t-1}\bigr).
\end{equation*}
Existing interactive retrievers update their queries mainly from the dialogue $\{(q_i,u_i)\}$. ReCoVR is \textit{closed-loop}: it also uses the ranking trajectory $\{\hat{L}_i\}_{i<t}$ to decide which candidates to demote and which result anchors the next composed query.

As shown in Fig.~\ref{fig:method}, each ReCoVR turn consists of three stages. The \textbf{Intent Pathway} decomposes feedback $u_t$ into structured, channel-specific queries for two fixed retrieval engines: a text-to-video (T2V) channel and a composed video retrieval (CoVR) channel. The \textbf{Reflection Pathway} uses feedback and semantic state to update query constraints. It also uses ranking history to lower the ranks of rejected or stagnant candidates. Finally, \textbf{Time-Weighted Reciprocal Rank Fusion (TW-RRF)} aggregates candidates from both channels over recent turns to produce the final ranked list $\hat{L}_t$.
ReCoVR uses two memory banks. \textbf{Progress Memory (PM)} stores semantic information. Its long-term part $\mathit{PM}^L$ caches gallery captions and metadata for reuse across sessions. Its short-term part $\mathit{PM}^S$ stores the running query and accumulated constraints for the current session. \textbf{Result Memory (RM)} stores ranked lists from both channels across turns. A vision-language model (VLM) supplies structured queries and correction signals rather than scoring videos directly, so ReCoVR runs on pre-trained retrievers without fine-tuning. Appendix~\ref{sec:appendix_algorithm} provides the turn-level algorithm. Appendix~\ref{sec:appendix_prompts} lists the prompts and output schemas used by every reasoning module.

\subsection{Reflexive Retrieval Mechanism}
\label{sec:dual_pathway}

The Intent Pathway (Sec.~\ref{sec:intent}) decides \textit{what to search for} from the latest feedback. The Reflection Pathway (Sec.~\ref{sec:reflection}) decides \textit{what to correct}: its constraints shape the next turn's queries, and rank-cap adjusts the current turn's lists before fusion. Both read the same session snapshot and can run in parallel; Algorithm~\ref{alg:recovr} in Appendix~\ref{sec:appendix_algorithm} gives the full order of operations in each turn.

\subsubsection{Intent Pathway}
\label{sec:intent}

The Intent Pathway translates ambiguous user feedback into executable queries (Fig.~\ref{fig:method}(b)). In multi-turn retrieval, user feedback often conveys mixed instructions. For instance, ``\textit{Make the keyboard black, add backlighting}'' mixes a relative edit with a target description; routing it to a single query type would discard part of the signal, so the \textbf{Intent Decomposer} maps $u_t$ into two queries:
\begin{equation}
  (\mathrm{search}_t,\; \mathrm{edit}_t)
  \;=\; \mathcal{I}(u_t,\; \mathit{PM}_{t-1}^S,\; \mathit{PM}^L).
  \label{eq:intent}
\end{equation}
Here, $\mathrm{search}_t$ is a target description synthesized from user feedback and accumulated context, while $\mathrm{edit}_t$ is a relative modification to the current result and may be empty for description-only feedback.

The \textbf{Query Reformer} appends the new intent to the running query in $\mathit{PM}_{t-1}^S$. It uses a large language model (LLM) to summarize text that exceeds a length threshold. This memory also contains constraints $\Delta c_{t-1}$ from the previous turn (with $\Delta c_0=\emptyset$), so the new query retains earlier corrections. Retrieval then proceeds through two channels:
\begin{equation}
  L_t^{T2V},\; L_t^{CoVR}
  \;=\; \mathcal{R}(\mathrm{search}_t,\; \mathrm{edit}_t,\;
                    \hat{L}_{t-1}(1),\; \mathcal{G}).
  \label{eq:retrieve}
\end{equation}

Here, $\hat{L}_{t-1}(1)$ denotes the top result shown to the user at the previous turn. The \textbf{T2V Channel} encodes $\mathrm{search}_t$ and retrieves videos by text-to-video similarity. The \textbf{CoVR Channel} uses $\mathrm{edit}_t$ to change the previously displayed result. When the edit is non-empty, it uses $\hat{L}_{t-1}(1)$ as the visual reference and retrieves by composed similarity. This dynamic anchor links a user's edit to the video they just saw. T2V thus supports direct target descriptions, while CoVR supports relative changes to the displayed video, so each form of feedback reaches the retriever suited to it.

An optional \textbf{Question Asking} module reads $\mathit{PM}_{t-1}^S$ and generates a clarifying question $q_{t+1}$. Prior I-TVR methods use questions and answers to refine queries~\citep{ivr_auto, merlin, umivr}. ReCoVR also accepts direct descriptions and edit instructions: users can answer the question or provide other feedback. Answering is never required: when the simulated user ignores the question entirely and follows a fixed modify-only or rewrite-only policy, ReCoVR still reaches 86.15\% and 92.14\% R@1 at Turn~5 (Appendix~\ref{sec:sim_robustness}).

\subsubsection{Reflection Pathway}
\label{sec:reflection}

The Reflection Pathway (Fig.~\ref{fig:method}(d)) combines semantic correction by the \textbf{State Reflector} with rank correction by the Rank-Cap Policy. The State Reflector evaluates user feedback $u_t$ in the context of the session state stored in $\mathit{PM}_{t-1}^S$, including accumulated queries, prior constraints, and gallery captions from $\mathit{PM}^L$. This joint analysis generates two structured outputs:
\begin{equation}
  (\sigma_t,\; \Delta c_t)
  \;=\; \mathcal{S}(u_t,\; \mathit{PM}_{t-1}^S,\; \mathit{PM}^L).
  \label{eq:reflect}
\end{equation}
We call this semantic update \textit{State Reflection}. The satisfaction level $\sigma_t \in \{\mathrm{negative},\, \mathrm{neutral}\}$ indicates whether the user explicitly rejects the current result or is providing incremental guidance. The constraint delta $\Delta c_t$ extracts semantic conditions based on the gap between user expectations and the current query state. It derives positive constraints that accumulate to guide ongoing searches, alongside negative constraints that act as semantic filters during query rewriting.

Beyond semantic constraints, ReCoVR needs an immediate mechanism to prevent clearly unhelpful results from dominating the next interaction. Inspired by negative-feedback reranking~\citep{wang2008study,bi2019conversational}, we instantiate a lightweight, failure-triggered \textbf{Rank-Cap Policy}:
\begin{equation}
  \Delta r_t \;=\; \mathcal{P}(\sigma_t,\; \mathit{RM}_{t-1}).
  \label{eq:rankcap}
\end{equation}
Let $\hat{v}=\hat{L}_{t-1}(1)$ be the previously displayed top result. The policy triggers on \textit{explicit rejection} or on \textit{strategy stagnation}, where the displayed top result persists across turns, and selects a cap position
\begin{equation*}
  \kappa_t =
  \begin{cases}
    \kappa_{\mathrm{neg}}, & \sigma_t=\mathrm{negative},\\
    \kappa_{\mathrm{neu}}, & \sigma_t=\mathrm{neutral}\ \text{and}\ \hat{L}_{t-1}(1)=\hat{L}_{t-2}(1),\\
    \infty, & \text{otherwise},
  \end{cases}
\end{equation*}
with $\kappa_{\mathrm{neu}}<\kappa_{\mathrm{neg}}$. In each channel list $L_t^{(s)}$, rank-cap moves $\hat{v}$ to rank $\max(r_{t,\hat{v}}^{(s)}, \kappa_t)$ and preserves the relative order of all other candidates; $\kappa_t=\infty$ leaves the lists unchanged. Capping rather than removing $\hat{v}$ keeps a partially relevant result available to later turns, and acting on ranks avoids comparing similarity scores that are not calibrated across channels (Appendix~\ref{sec:rank_fuse}); stagnation receives the milder cap because a persisting result is weaker evidence than an explicit rejection. Demoting $\hat{v}$ also keeps an unreliable result from becoming the visual anchor of the next turn.

\subsection{Time-Weighted Reciprocal Rank Fusion}
\label{sec:twrrf}

As interaction progresses, ReCoVR collects ranked lists from both T2V and CoVR channels across turns. Standard Reciprocal Rank Fusion (RRF)~\citep{rrf} combines multiple ranked lists by rewarding candidates that appear consistently at high ranks. However, directly applying RRF to all historical turns ignores the temporal nature of ICoVR: later feedback usually reflects a more refined intent, while early turns may encode ambiguous or outdated constraints. We therefore extend RRF into a temporally aware fusion rule by adding recency weighting and a sliding temporal window, yielding \textbf{Time-Weighted Reciprocal Rank Fusion (TW-RRF)}.

For current turn $t$ and window size $W$, TW-RRF aggregates $N=t-t_0+1$ turns, from $t_0=\max(1,t-W+1)$ to $t$. The normalized linear recency weight for turn $t'$ is:
\begin{equation}
  w_{t'} \;=\; \frac{2(t' - t_0 + 1)}{N(N+1)}.
  \label{eq:tw_weight}
\end{equation}
Recent turns thus contribute more, and the window keeps early ambiguous feedback from dominating after the user has refined the intent.
For a candidate video $v_i$, let $r_{t',i}^{(s)}$ be its rank in channel $s \in \{\text{T2V},\text{CoVR}\}$ at turn $t'$, and let $\mathbbm{1}[v_i \in L_{t'}^{(s)}]$ indicate whether it appears in the top-$K$ list. The final list $\hat{L}_t$ is obtained by sorting candidates according to:
\begin{equation}
  \hat{L}_t \;=\; \underset{v_i}{\mathrm{sort}^{\downarrow}}
  \Bigg[\,
    \sum_{s \in \{\text{T2V},\text{CoVR}\}}
    \sum_{t'=t_0}^{t}
    w_{t'} \cdot
    \frac{\mathbbm{1}[v_i \in L_{t'}^{(s)}]}{k + r_{t',i}^{(s)}}
  \,\Bigg],
  \label{eq:twrrf}
\end{equation}
where $k$ is the standard RRF smoothing constant. When all turns are assigned equal weights and the window covers the full history, TW-RRF reduces to conventional RRF over multi-turn channel outputs.
TW-RRF operates on the channel lists after rank-cap (Eq.~(\ref{eq:rankcap})): the current lists are adjusted before being stored in RM and fused, earlier lists are not rewritten, and an adjusted list influences later turns until it leaves the window (Appendix~\ref{sec:appendix_algorithm}).

\section{Experiments}
\label{sec:exp}

\subsection{Experimental Setup}
\label{sec:setup_exp}

\textbf{Datasets and Interactive Baselines.}
We evaluate on \textbf{WebVid-CoVR}~\citep{webvid_covr} (2,556 manually annotated triplets), \textbf{Dense-WebVid-CoVR}~\citep{densewebvidcovr} (the same video pairs with richer modification texts: 31.16 vs.\ 4.6 words on average), and \textbf{FineCVR}~\citep{finecvr} (10,043 fine-grained temporal queries over a larger, more diverse gallery).
Generalization benchmarks include \textbf{FashionIQ}~\citep{fashioniq} for composed image retrieval (three clothing categories), and the text-to-video benchmarks \textbf{MSRVTT-1kA}~\citep{xu2016msr-vtt, yu2018joint}, and \textbf{AVSD-1k-Test}~\citep{avsd} (1,000 queries each for the latter two, following prior protocols~\citep{d2v,vired,umivr}).
We adapt \texttt{IVR}~\citep{ivr_auto}, \texttt{Merlin}~\citep{merlin}, and \texttt{UMIVR}~\citep{umivr} as interactive baselines, equipping each with the same single-turn CoVR retriever as ReCoVR.

\textbf{User Simulator.}
Following simulated-user evaluation in interactive retrieval~\citep{zhang2024usimagent,balog2023user,ivr_auto,umivr,chen2026adept, dou2025simulatorarena}, a VLM compares the target video with the current top-ranked candidate to generate feedback. It serves solely as an \textit{evaluation instrument}: ReCoVR and the baselines never access the target, receiving only feedback and their own retrieval history. All methods share the simulator, prompts, targets, initial rankings, and stopping criteria. The simulator chooses between a relative edit and a fresh target description from how far the displayed result is from the target, independently of the method under evaluation, and replacing ReCoVR's reasoning VLM with InternVL3.5 while keeping the simulator fixed preserves the gains (Appendix~\ref{sec:reasoning_vlm}). Appendix~\ref{sec:supp_user_sim} details the simulator and its prompts, and Sec.~\ref{sec:user_study} reports evaluation with real human feedback.

\textbf{Evaluation Protocol.}
Starting from a CoVR query at Turn~0, each interaction allows up to $T{=}5$ feedback turns, presenting the top-ranked result before each update. We report per-turn Recall at $K$ (R@$K$) and Best log Rank Integral (BRI)~\citep{plugir}. BRI integrates the logarithm of the best target rank attained so far, rewarding early retrieval success across the trajectory; lower is better.

\textbf{Implementation Details.}
Unless specified otherwise, ReCoVR and the simulator use Qwen3-VL-4B~\citep{Qwen3-VL}, and both retrieval channels use BLIP-ViT-Large~\citep{li2022blip}. We set the TW-RRF window $W{=}5$, RRF constant $k{=}60$, candidate pool $K{=}100$, and rank-cap positions $\kappa_{\mathrm{neg}}{=}11$ and $\kappa_{\mathrm{neu}}{=}4$. All models remain frozen; ReCoVR requires no training.

\begin{table*}[tbp]
\begin{minipage}[b]{0.515\textwidth}
    \caption{\textbf{Comparison with interactive baselines} on WebVid-CoVR-Test. ReCoVR leads in R@1 at every feedback turn, with the margin over the strongest baseline growing from 7.2 to 15.2 points, and it matches or exceeds all baselines on R@5, R@10, and BRI at every turn.}
    \label{tab:compare_covr}
    \centering
    \resizebox{\linewidth}{!}{
        \begin{tabular}{llcccccc}
        \toprule[1pt]
                                  & Methods      & \textit{Turn 0} & \textit{Turn 1}          & \textit{Turn 2}          & \textit{Turn 3}          & \textit{Turn 4}          & \textit{Turn 5}         \\ \hline
        \multirow{4}{*}{\rotatebox{90}{\textbf{R@1} $\uparrow$}} & IVR          & 54.30  & 66.55           & 70.11           & 71.95           & 73.00           & 73.51           \\
                                  & Merlin       & 54.30  & 60.80           & 63.34           & 64.63           & 65.49           & 65.77           \\
                                  & UMIVR        & 54.30  & 67.10           & 71.09           & 72.85           & 74.10           & 75.04           \\
                            &  \cellcolor{tbgreen!70}ReCoVR(ours) & \cellcolor{tbgreen!70}54.30  & \cellcolor{tbgreen!70}\textbf{74.30}  & \cellcolor{tbgreen!70}\textbf{80.32}  & \cellcolor{tbgreen!70}\textbf{84.39}  & \cellcolor{tbgreen!70}\textbf{87.72}  & \cellcolor{tbgreen!70}\textbf{90.22}  \\ \hline
        \multirow{4}{*}{\rotatebox{90}{\textbf{R@5} $\uparrow$}} & IVR          & 77.00  & 85.45           & 87.68           & 88.93           & 89.75           & 90.10           \\
                                  & Merlin       & 77.00  & 79.23           & 82.75           & 83.92           & 84.43           & 84.62           \\
                                  & UMIVR        & 77.00  & 86.62           & \textbf{89.28}  & 90.02           & 90.18           & 90.41           \\
                                  & \cellcolor{tbgreen!70}ReCoVR(ours) & \cellcolor{tbgreen!70}77.00  & \cellcolor{tbgreen!70}\textbf{87.28}  & \cellcolor{tbgreen!70}\textbf{89.28}  & \cellcolor{tbgreen!70}\textbf{91.28}  & \cellcolor{tbgreen!70}\textbf{93.04}  & \cellcolor{tbgreen!70}\textbf{94.13}  \\ \hline
        \multirow{4}{*}{\rotatebox{90}{\textbf{R@10} $\uparrow$}} & IVR          & 84.35  & 91.08           & 93.78           & 94.44           & 94.84           & 95.03          \\
                                  & Merlin       & 84.35  & 86.27           & 89.63          & 90.49           & 91.20           & 91.47           \\
                                  & UMIVR        & 84.35  & 92.64           & 93.56  & 94.33           & 94.56           & 94.99           \\
                                  & \cellcolor{tbgreen!70}ReCoVR(ours) & \cellcolor{tbgreen!70}84.35  & \cellcolor{tbgreen!70}\textbf{93.08}  & \cellcolor{tbgreen!70}\textbf{93.81}  & \cellcolor{tbgreen!70}\textbf{94.56}  & \cellcolor{tbgreen!70}\textbf{95.27}  & \cellcolor{tbgreen!70}\textbf{96.24}  \\ \hline
        \multirow{4}{*}{\rotatebox{90}{\textbf{BRI} $\downarrow$}}      & IVR          & -      & 0.7368          & 0.6187          & 0.5560          & 0.5161          & 0.4880          \\
                                  & Merlin       & -      & 0.8165          & 0.7399          & 0.6921          & 0.6603          & 0.6381          \\
                                  & UMIVR        & -      & 0.7285          & 0.6049          & 0.5394          & 0.4986          & 0.4700          \\
                            & \cellcolor{tbgreen!70}ReCoVR(ours) & \cellcolor{tbgreen!70}-      & \cellcolor{tbgreen!70}\textbf{0.6737} & \cellcolor{tbgreen!70}\textbf{0.5271} & \cellcolor{tbgreen!70}\textbf{0.4538} & \cellcolor{tbgreen!70}\textbf{0.4034} & \cellcolor{tbgreen!70}\textbf{0.3643} \\ \bottomrule[1pt]
        \end{tabular}
    }
\par\vspace{0pt}
\end{minipage}
\hfill
\begin{minipage}[b]{0.455\textwidth}
    \caption{\textbf{Reference comparison with single-turn CoVR methods} on WebVid-CoVR-Test. Interactive turns use extra feedback, so this is not a like-for-like benchmark. \textbf{Bold} marks the first turn at which ReCoVR matches the best single-turn result.}
    \label{tab:compare_oneturn}
    \centering
    \resizebox{\linewidth}{!}{
        \begin{tabular}{lccccc}
        \toprule[1pt]
        \diagbox{Methods}{Metrics}     & \textbf{R@1} $\uparrow$   & \textbf{R@5} $\uparrow$  & \textbf{R@10} $\uparrow$  & \textbf{R@50} $\uparrow$ & \textbf{Avg.} $\uparrow$  \\ \hline
        \rowcolor{tbgrey!70}\multicolumn{6}{c}{\textit{Single-Turn CoVR}}                                                        \\ \hline
        CoVR$_{AAAI~24}$         & 53.13          & 79.93          & 86.85          & 97.69          & 79.40          \\
        ECDE$_{CVPR~24}$   & 60.12          & 84.32          & 91.27          & 98.72          & 83.61          \\
        CoVR-2$_{TPAMI~24}$       & 59.82          & 83.84          & 91.28          & 98.24          & 83.30          \\
        TFR-CVR$_{ECCV~24}$         & 51.70          & 75.30          & 80.70          & -              & -              \\
        FDCA$_{ICLR~25}$           & 54.80          & 82.27          & 89.84          & 97.70          & 81.15          \\
        HUD$_{MM~25}$        & 63.38          & 86.93          & 92.29          & 98.76          & 85.34          \\
        CADP$_{PR~25}$        & 63.62          & 87.36          & 93.19          & 99.30          & 85.87          \\
        REFINE$_{TOMM~26}$        & 62.48          & 86.82          & 92.41          & 98.87          & 85.15          \\ \hline
        \rowcolor{tbgreen!70}\multicolumn{6}{c}{\textit{ReCoVR (Our Interactive CoVR)}}                                           \\ \hline
        \multicolumn{1}{c}{\textit{Turn 0}} & 54.30          & 77.00          & 84.35          & 95.77          & 77.86          \\
        \multicolumn{1}{c}{\textit{Turn 1}} & \textbf{74.30} & 87.28          & 93.08          & 98.75          & \textbf{88.35} \\
        \multicolumn{1}{c}{\textit{Turn 2}} & 80.32          & \textbf{89.28} & \textbf{93.81} & 98.90          & 90.45          \\
        \multicolumn{1}{c}{\textit{Turn 3}} & 84.39          & 91.28          & 94.56          & 99.18          & 92.35          \\
        \multicolumn{1}{c}{\textit{Turn 4}} & 87.72          & 93.04          & 95.27          & \textbf{99.30} & 93.83          \\
        \multicolumn{1}{c}{\textit{Turn 5}} & 90.22          & 94.13          & 96.24          & 99.57          & 95.04          \\ \bottomrule[1pt]
        \end{tabular}
    }
\par\vspace{0pt}
\end{minipage}
\end{table*}

\subsection{Main Comparison under the ICoVR Protocol}
\label{sec:main_results}

Table~\ref{tab:compare_covr} compares interactive methods under the shared ICoVR protocol. ReCoVR leads in R@1 at every feedback turn, and its margin over the strongest baseline grows across turns. Paired session-level 95\% confidence intervals for the per-turn R@1 advantage lie above zero (Appendix~\ref{sec:paired_ci}).

The trajectory analysis in Appendix~\ref{sec:failure_diag} explains why the margin grows. UMIVR fails to improve the target rank in more than half of its active feedback turns, far more often than ReCoVR, and its failures tend to persist: after a failed turn, the rank improves on the next turn in only 30.12\% of cases, below UMIVR's own average. ReCoVR recovers in 88.15\% of cases, above its own average. An open-loop system that stalls tends to stay stalled, whereas ReCoVR registers the failure in its ranking history and corrects it. These rates are computed offline on each method's own trajectories; target ranks are never available to the system during search.

\begin{figure*}[tbp]
    \centering
    \includegraphics[width=1.0\linewidth]{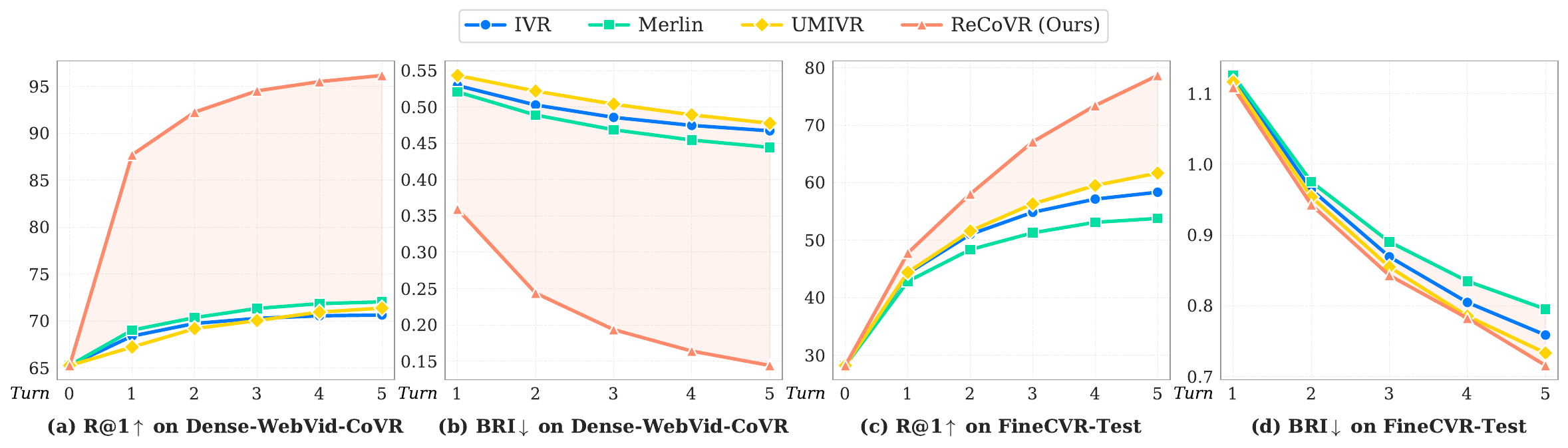}
    \vspace{-2pt}
    \caption{\textbf{ICoVR comparison on additional CoVR benchmarks.}
Under the same interactive protocol, ReCoVR shows stronger turn-wise improvement than adapted I-TVR baselines on Dense-WebVid-CoVR-Test (a,b) and FineCVR-Test (c,d). The shaded region marks ReCoVR's margin over the baselines, which widens across turns in R@1.}
    \label{fig:more_compare}
    \vspace{-3pt}
\end{figure*}

Fig.~\ref{fig:more_compare} shows the same pattern on Dense-WebVid-CoVR and FineCVR. Dense-WebVid-CoVR pairs the same videos with much longer modification texts, so the continued gains show that interaction helps even when the initial query is already detailed.

Table~\ref{tab:compare_oneturn} compares against single-turn CoVR~\citep{webvid_covr}, ECDE~\citep{covr_enrich}, CoVR-2~\citep{webvid_covr2}, TFR-CVR~\citep{egocvr}, FDCA~\citep{finecvr}, HUD~\citep{hud}, CADP~\citep{cadp}, and REFINE~\citep{refine}. Starting from a mid-range backbone, ReCoVR surpasses the best single-turn method in R@1 after a single feedback turn and keeps improving with further interaction.

\begin{table*}[tbp]
    \caption{\textbf{Component ablation} on WebVid-CoVR-Test and FineCVR-Test. Each row enables the checked components with all other settings fixed; without State Reflection, the satisfaction signal $\sigma_t$ is set to neutral. Dual-channel routing, State Reflection, and rank-cap each contribute, and the full model has the best R@1 at every turn on both benchmarks.}
    \label{tab:ablation}
    \begin{center}
    \resizebox{0.999\linewidth}{!}{
\begin{tabular}{ccccc|l|lllll|lllll}
\toprule[1pt]
\multirow{2}{*}{\textit{\begin{tabular}[c]{@{}c@{}}CoVR \\ Channel\end{tabular}}} & \multirow{2}{*}{\textit{\begin{tabular}[c]{@{}c@{}}T2V \\ Channel\end{tabular}}} & \multirow{2}{*}{\textit{\begin{tabular}[c]{@{}c@{}}Intent \\ Pathway\end{tabular}}} & \multirow{2}{*}{\textit{\begin{tabular}[c]{@{}c@{}}State \\ Reflection\end{tabular}}} & \multirow{2}{*}{\textit{\begin{tabular}[c]{@{}c@{}}Rank-\\ cap\end{tabular}}} & \multicolumn{1}{c|}{\textbf{Dataset}} & \multicolumn{5}{c|}{\textbf{WebVid-CoVR-Test}}                                                                                                                           & \multicolumn{5}{c}{\textbf{FineCVR-Test}}                                                                                                                               \\
                                                                                  &                                                                                  &                                                                                     &                                                                                       &                                                                               & \multicolumn{1}{c|}{\textbf{Metrics}} & \multicolumn{1}{c}{\textit{Turn 1}} & \multicolumn{1}{c}{\textit{2}} & \multicolumn{1}{c}{\textit{3}} & \multicolumn{1}{c}{\textit{4}} & \multicolumn{1}{c|}{\textit{5}} & \multicolumn{1}{c}{\textit{Turn 1}} & \multicolumn{1}{c}{\textit{2}} & \multicolumn{1}{c}{\textit{3}} & \multicolumn{1}{c}{\textit{4}} & \multicolumn{1}{c}{\textit{5}} \\ \hline
\multirow{2}{*}{$\checkmark$}                                                                & \multirow{2}{*}{}                                                                & \multirow{2}{*}{}                                                                   & \multirow{2}{*}{}                                                                     & \multirow{2}{*}{}                                                             & R@1$\uparrow$                                 & \multicolumn{1}{c}{68.15}           & \multicolumn{1}{c}{71.64}      & \multicolumn{1}{c}{72.73}      & \multicolumn{1}{c}{73.40}      & \multicolumn{1}{c|}{74.52}      & \multicolumn{1}{c}{44.54}           & \multicolumn{1}{c}{51.92}      & \multicolumn{1}{c}{56.28}      & \multicolumn{1}{c}{59.60}      & \multicolumn{1}{c}{61.34}      \\
                                                                                  &                                                                                  &                                                                                     &                                                                                       &                                                                               & BRI$\downarrow$                                   & \multicolumn{1}{c}{0.6938}          & \multicolumn{1}{c}{0.5699}     & \multicolumn{1}{c}{0.5192}     & \multicolumn{1}{c}{0.4890}     & \multicolumn{1}{c|}{0.4679}     & \multicolumn{1}{c}{1.1249}          & \multicolumn{1}{c}{0.9576}     & \multicolumn{1}{c}{0.8607}     & \multicolumn{1}{c}{0.7895}     & \multicolumn{1}{c}{0.7487}     \\ \cline{6-16} 
\multirow{2}{*}{}                                                                 & \multirow{2}{*}{$\checkmark$}                                                               & \multirow{2}{*}{}                                                                   & \multirow{2}{*}{}                                                                     & \multirow{2}{*}{}                                                             & R@1$\uparrow$                                & \multicolumn{1}{c}{66.76}           & \multicolumn{1}{c}{71.67}      & \multicolumn{1}{c}{72.42}      & \multicolumn{1}{c}{73.08}      & \multicolumn{1}{c|}{73.59}      & \multicolumn{1}{c}{43.39}           & \multicolumn{1}{c}{49.70}      & \multicolumn{1}{c}{54.21}      & \multicolumn{1}{c}{58.56}      & \multicolumn{1}{c}{60.55}      \\
                                                                                  &                                                                                  &                                                                                     &                                                                                       &                                                                               & BRI$\downarrow$                                   & \multicolumn{1}{c}{0.7061}          & \multicolumn{1}{c}{0.5724}     & \multicolumn{1}{c}{0.5210}     & \multicolumn{1}{c}{0.4906}     & \multicolumn{1}{c|}{0.4694}     & \multicolumn{1}{c}{1.1865}          & \multicolumn{1}{c}{0.9579}     & \multicolumn{1}{c}{0.8569}     & \multicolumn{1}{c}{0.7962}     & \multicolumn{1}{c}{0.7507}     \\ \cline{6-16} 
\multirow{2}{*}{$\checkmark$}                                                                & \multirow{2}{*}{$\checkmark$}                                                               & \multirow{2}{*}{$\checkmark$}                                                                  & \multirow{2}{*}{}                                                                     & \multirow{2}{*}{}                                                             & R@1$\uparrow$                              & 69.29                               & 72.69                          & 74.61                          & 76.17                          & 77.46                           & 45.56                               & 52.17                          & 58.85                          & 61.22                          & 63.69                          \\
                                                                                  &                                                                                  &                                                                                     &                                                                                       &                                                                               & BRI$\downarrow$                                   & 0.6915                              & 0.5619                         & 0.5017                         & 0.4733                         & 0.4590                          & 1.1151                              & 0.9556                         & 0.8518                         & 0.7859                         & 0.7310                         \\ \cline{6-16} 
\multirow{2}{*}{$\checkmark$}                                                                &                                                               &                                                                   & \multirow{2}{*}{$\checkmark$}                                                                      & \multirow{2}{*}{}                                                             & R@1$\uparrow$                              & 68.94                               & 73.63                         & 75.36                          & 76.28                         & 78.60                           & 45.42                               & 52.71                          & 57.94                          & 62.23                         & 65.38                          \\
                                                                                  &                                                                                  &                                                                                     &                                                                                       &                                                                               & BRI$\downarrow$                                   & 0.6935                              & 0.5608                         & 0.5114                         & 0.4860                         & 0.4533                          & 1.1243                              & 0.9556                         & 0.8584                         & 0.7897                         & 0.7486                         \\ \cline{6-16}
\multirow{2}{*}{$\checkmark$}                                                                &                                                               &                                                                   & \multirow{2}{*}{$\checkmark$}                                                                      & \multirow{2}{*}{$\checkmark$}                                                             & R@1$\uparrow$                              & 71.61                               & 78.11                         & 81.46                          & 83.65                         & 85.17                           & 46.85                               & 56.13                          & 62.34                          & 69.97                         & 73.89                          \\
                                                                                  &                                                                                  &                                                                                     &                                                                                       &                                                                               & BRI$\downarrow$                                   & 0.6804                              & 0.5445                         & 0.4832                         & 0.4330                         & 0.4019                          & 1.1204                              & 0.9553                         & 0.8544                         & 0.7872                         & 0.7380                         \\ \cline{6-16}
\multirow{2}{*}{$\checkmark$}                                                                & \multirow{2}{*}{$\checkmark$}                                                               & \multirow{2}{*}{$\checkmark$}                                                                  & \multirow{2}{*}{$\checkmark$}                                                                    & \multirow{2}{*}{}                                                             & R@1$\uparrow$                                   & 70.42                               & 75.51                          & 78.44                          & 80.13                          & 83.26                           & 46.84                               & 53.24                          & 61.14                          & 64.95                          & 69.40                          \\
                                                                                  &                                                                                  &                                                                                     &                                                                                       &                                                                               & BRI$\downarrow$                                   & 0.6918                              & 0.5491                         & 0.4912                         & 0.4691                         & 0.4377                          & 1.1143                              & 0.9527                         & 0.8486                         & 0.7858                         & 0.7284                         \\ \cline{6-16} 
\multirow{2}{*}{$\checkmark$}                                                                & \multirow{2}{*}{$\checkmark$}                                                               & \multirow{2}{*}{$\checkmark$}                                                                  & \multirow{2}{*}{}                                                                     & \multirow{2}{*}{$\checkmark$}                                                            & R@1$\uparrow$                                   & 72.41                               & 79.39                          & 83.94                          & 86.70                          & 87.52                           & 47.48                               & 55.66                          & 64.17                          & 71.85                          & 76.87                          \\
                                                                                  &                                                                                  &                                                                                     &                                                                                       &                                                                               & BRI$\downarrow$                                   & 0.6810                              & 0.5355                         & 0.4847                         & 0.4241                         & 0.3964                          & 1.1140                              & 0.9531                         & 0.8453                         & 0.7841                         & 0.7171                         \\ \cline{6-16} 
\multirow{2}{*}{$\checkmark$}                                                                & \multirow{2}{*}{$\checkmark$}                                                               & \multirow{2}{*}{$\checkmark$}                                                                  & \multirow{2}{*}{$\checkmark$}                                                                    & \multirow{2}{*}{$\checkmark$}                                                            & R@1$\uparrow$                                   & 74.30                               & 80.32                          & 84.39                          & 87.72                          & 90.22                           & 47.78                               & 58.04                          & 67.11                          & 73.42                          & 78.65                          \\
                                                                                  &                                                                                  &                                                                                     &                                                                                       &                                                                               & BRI$\downarrow$                                   & 0.6737                              & 0.5271                         & 0.4538                         & 0.4034                         & 0.3643                          & 1.1087                              & 0.9529                         & 0.8430                         & 0.7825                         & 0.7157                         \\ \bottomrule[1pt]
\end{tabular}

    }
\end{center}
\end{table*}

\subsection{Ablation Study}
\label{sec:ablation}

\textbf{Component contributions.} Table~\ref{tab:ablation} adds components one at a time, and three trends emerge. First, the two channels are complementary: the CoVR-only and T2V-only variants perform similarly, yet routing feedback to both through the Intent Pathway improves over either alone, so each channel succeeds on feedback that the other handles poorly. Second, the two correction mechanisms add up. Rank-cap gives the larger single gain on both benchmarks, but without State Reflection $\sigma_t$ defaults to neutral and rank-cap can only react to stagnation. Adding State Reflection supplies rejection signals and improves R@1 and BRI at every turn, and the gap to the Intent-only variant widens as the interaction proceeds. Rank-cap corrects the current ranking immediately, while State Reflection shapes later queries. Third, the gains do not come from the extra retriever: with the CoVR channel alone, State Reflection and rank-cap already surpass every adapted baseline at every feedback turn.

\textbf{Simpler uses of history.} A baseline that regenerates one composed query from the full interaction history, without reflection, improves over turns but falls increasingly behind ReCoVR, and its R@50 even drops below the Turn-0 level. Re-deriving a single query from a growing history thus tends to over-condition the ranking. Transplanting rank-cap into IVR and UMIVR helps both, yet their gap to ReCoVR still widens over turns (Appendix~\ref{sec:rankcap}). Appendix~\ref{sec:fullhistory} compares further single-mechanism strategies, all of which trail the full loop.

\textbf{Fusion and anchoring.} TW-RRF gives the best final retrieval among the fusion variants (Appendix~\ref{sec:rank_fuse}). Uniform turn weights come close early but fall behind in later turns, current-turn-only fusion is competitive only at Turn~1, and score-level fusion is weaker throughout because raw similarities are not comparable across channels. Cross-turn evidence and recency weighting therefore both matter. Dynamic anchoring outperforms a fixed reference at every turn, since each edit then refers to the video the user has just seen (Appendix~\ref{sec:anchor}).

\begin{figure*}[tbp]
    \centering
    \includegraphics[width=0.9\linewidth]{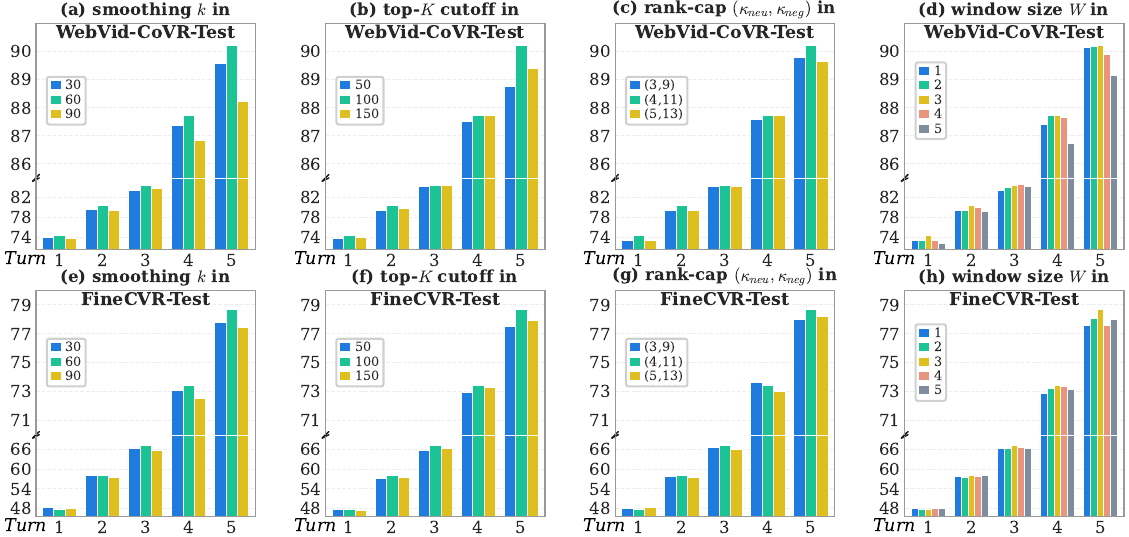}
    \caption{
\textbf{Parameter sensitivity of TW-RRF} on WebVid-CoVR-Test (top) and FineCVR-Test (bottom). ReCoVR remains stable across broad ranges of $k$, top-$K$ cutoff, rank-cap positions, and window size $W$, indicating that its gains do not depend on precise parameter tuning.
}
\label{fig:ab_param}
\end{figure*}

\textbf{Sensitivity.} Performance is stable across the tested $k$, $K$, and rank-cap positions (Fig.~\ref{fig:ab_param}), and a larger window $W$ helps until gains saturate, indicating that recent history is useful while much older turns add little. Rewording all module prompts barely changes final accuracy, and nearly all structured outputs parse on the first attempt (Appendix~\ref{sec:prompt_robust}).

\textbf{Generalization.} Appendix~\ref{sec:generalization} shows gains on FashionIQ, across nine backbone configurations and two reasoning-VLM families, and on I-TVR (MSRVTT, AVSD) when reflection is added to UMIVR under its original protocol, so reflection also helps existing pipelines.
\begin{table*}[tbp]
    \caption{\textbf{Human versus simulated feedback on WebVid-CoVR-Test.} On 100 ICoVR instances, ReCoVR follows a closely matching improvement trend under free-form human feedback and simulated feedback. IVR with simulated feedback is shown for reference.}
    \vspace{-0.4em}
    \label{tab:user_study}
    \begin{center}
    \resizebox{0.9\linewidth}{!}{
\begin{tabular}{lc|cccc|cccc|ccc|c}
\toprule[1pt]
\multirow{2}{*}{Method} & \multirow{2}{*}{\begin{tabular}[c]{@{}l@{}}User\\ Type\end{tabular}} & \multicolumn{4}{c|}{\textbf{R@1$\uparrow$}}                      & \multicolumn{4}{c|}{\textbf{R@5$\uparrow$}}                 & \multicolumn{3}{c|}{\textbf{BRI$\downarrow$}}    & \multirow{2}{*}{\begin{tabular}[c]{@{}c@{}}Feedback\\ Words(Avg.)\end{tabular}} \\
                        &                                                                      & \textit{Turn 0} & \textit{1} & \textit{2} & \textit{3} & \textit{0} & \textit{1} & \textit{2} & \textit{3} & \textit{1} & \textit{2} & \textit{3} &                                                                                 \\ \hline
IVR                     & Simulator                                                            & 43.0            & 58.0       & 58.0       & 59.0       & 63.0       & 75.0       & 82.0       & 81.0       & 1.005      & 0.860      & 0.788      & 6.1                                                                             \\
ReCoVR                  & Simulator                                                            & 43.0            & 63.0       & 73.0       & 79.0       & 63.0       & 83.0       & 83.0       & 85.0       & 0.943      & 0.751      & 0.650      & 6.3                                                                             \\
ReCoVR                  & Real Humans                                                          & 43.0            & 62.0       & 70.0       & 76.0       & 63.0       & 77.0       & 80.0       & 84.0       & 0.977      & 0.800      & 0.700      & 13.7                                                                            \\ \bottomrule[1pt]
\end{tabular}
    }
\end{center}
\end{table*}

\subsection{Human Feedback and Computational Cost}
\label{sec:user_study}

Simulated feedback is reproducible~\citep{zhang2024usimagent,ivr_auto,chatir,umivr} but follows a fixed policy, so we check whether it reflects real interaction by running ReCoVR with free-form feedback from real users, following the protocol of~\citet{ivr_auto}. Five participants interact with ReCoVR on 100 WebVid-CoVR-Test instances for up to three turns each, through a web interface that displays the target video, the current top-ranked result, and any system question (Appendix~\ref{sec:supp_user_interface}). Table~\ref{tab:user_study} shows closely matching trajectories under human and simulated feedback, both well above IVR with simulated feedback. Human feedback is roughly twice as long and less templated than simulated feedback, yet ReCoVR turns it into comparable gains (protocol in Appendix~\ref{sec:supp_user_sim}). ReCoVR also retains its advantage under simulated feedback with noisy wording, omitted constraints, or modify-only edits (Appendix~\ref{sec:sim_robustness}).
\textbf{Computational Cost.}\label{sec:analysis}
On 100 WebVid-CoVR-Test instances with one RTX A6000 GPU (Appendix~\ref{sec:appendix_efficiency}), ReCoVR needs fewer feedback turns than UMIVR but more time per turn. VLM reasoning dominates this cost, while retrieval and rank fusion take only hundredths of a second per turn and the memory bank adds a few megabytes. Under a conservative estimate of human response time, ReCoVR's own computation is about 15\% of end-to-end latency, so the user rather than the model sets the pace of practical interaction, and a smaller or faster reasoning model would reduce it further.

\section{Conclusion}
\label{sec:conclusion}

We introduced ICoVR and showed that adapted interactive retrievers often stall because they ignore their own ranking history. ReCoVR closes this loop without training and improves over these baselines on three CoVR benchmarks, with a margin that grows across turns. More broadly, a retriever's own ranking history is an inexpensive and underused signal: it needs no access to the target and no retraining, yet it identifies the turns at which feedback alone fails. Next steps include adaptive channel weighting for large reference--target gaps (Appendix~\ref{sec:supp_case_study}), lighter reasoning models for lower per-turn cost, and user studies with more diverse participants and search behaviors.

\ifarxiv\else\clearpage\fi
\section*{Reproducibility Statement}
The experimental setup is described in Section~\ref{sec:setup_exp}. Appendix~\ref{sec:appendix_algorithm} provides the turn-level algorithm, Appendix~\ref{sec:appendix_prompts} lists the prompts and output schemas, and Appendix~\ref{sec:supp_user_sim} describes the user simulation and human-feedback protocols. Additional analyses and computational details are provided in Appendices~\ref{sec:appendix_ablation} and~\ref{sec:appendix_efficiency}. We will release source code in the supplementary material to support reproduction.

\section*{Ethics Statement}
The human-feedback study received institutional ethics approval. Participants were informed of the task procedure and minimal risks, and no demographic attributes or sensitive personal information were collected. Further details are provided in Appendix~\ref{sec:supp_user_interface}. Video retrieval over sensitive or personal collections may raise privacy concerns; responsible deployment should consider data consent, access control, and dataset and model biases, as discussed in Appendix~\ref{sec:appendix_limitation}.

\section*{AI Use Statement}
\label{sec:ai_use}
LLMs assisted with manuscript drafting and language polishing, as well as the implementation and checking of parts of the code. The authors reviewed the AI-assisted work and remain responsible for the accuracy and integrity of the manuscript and code.

\bibliography{iclr2027_conference}
\bibliographystyle{iclr2027_conference}

\appendix
\clearpage

\let\AND\relax

\begingroup
\hypersetup{linkcolor=appendixteal}
\color{appendixteal}
\small
\noindent
\hyperref[sec:appendix_ablation]{\textbf{A\hspace{0.8em}Additional Analyses and Ablations}}\dotfill\pageref{sec:appendix_ablation}\par\vspace{1pt}
\hspace{1.2em}\hyperref[sec:rank_fuse]{A.1\hspace{0.6em}Rank Fusion Strategy}\dotfill\pageref{sec:rank_fuse}\par\vspace{1pt}
\hspace{1.2em}\hyperref[sec:rankcap]{A.2\hspace{0.6em}Baseline Rank-cap Transfer}\dotfill\pageref{sec:rankcap}\par\vspace{1pt}
\hspace{1.2em}\hyperref[sec:paired_ci]{A.3\hspace{0.6em}Statistical Significance via Paired Bootstrap}\dotfill\pageref{sec:paired_ci}\par\vspace{1pt}
\hspace{1.2em}\hyperref[sec:failure_diag]{A.4\hspace{0.6em}Measuring Open-Loop Failure Modes}\dotfill\pageref{sec:failure_diag}\par\vspace{1pt}
\hspace{1.2em}\hyperref[sec:prompt_robust]{A.5\hspace{0.6em}Robustness to Prompt Rewording and Structured-Output Reliability}\dotfill\pageref{sec:prompt_robust}\par\vspace{1pt}
\hspace{1.2em}\hyperref[sec:anchor]{A.6\hspace{0.6em}Fixed-Reference versus Dynamic Anchoring}\dotfill\pageref{sec:anchor}\par\vspace{1pt}
\hspace{1.2em}\hyperref[sec:fullhistory]{A.7\hspace{0.6em}Simplified Multi-Turn Strategies}\dotfill\pageref{sec:fullhistory}\par\vspace{2pt}
\noindent
\hyperref[sec:generalization]{\textbf{B\hspace{0.8em}Generalization}}\dotfill\pageref{sec:generalization}\par\vspace{2pt}
\noindent
\hyperref[sec:appendix_efficiency]{\textbf{C\hspace{0.8em}Efficiency Analysis}}\dotfill\pageref{sec:appendix_efficiency}\par\vspace{2pt}
\noindent
\hyperref[sec:supp_user_sim]{\textbf{D\hspace{0.8em}User Simulation and Human-Feedback Study Details}}\dotfill\pageref{sec:supp_user_sim}\par\vspace{1pt}
\hspace{1.2em}\hyperref[sec:supp_sim_motivation]{D.1\hspace{0.6em}Motivation: Why Simulate Users?}\dotfill\pageref{sec:supp_sim_motivation}\par\vspace{1pt}
\hspace{1.2em}\hyperref[sec:supp_sim_design]{D.2\hspace{0.6em}User Simulator Design}\dotfill\pageref{sec:supp_sim_design}\par\vspace{1pt}
\hspace{1.2em}\hyperref[sec:supp_user_interface]{D.3\hspace{0.6em}Human-Feedback Interface}\dotfill\pageref{sec:supp_user_interface}\par\vspace{1pt}
\hspace{1.2em}\hyperref[sec:sim_robustness]{D.4\hspace{0.6em}Robustness to Degraded Simulated Feedback}\dotfill\pageref{sec:sim_robustness}\par\vspace{2pt}
\noindent
\hyperref[sec:supp_case_study]{\textbf{E\hspace{0.8em}Qualitative Case Studies}}\dotfill\pageref{sec:supp_case_study}\par\vspace{2pt}
\noindent
\hyperref[sec:appendix_algorithm]{\textbf{F\hspace{0.8em}Algorithmic Overview of ReCoVR}}\dotfill\pageref{sec:appendix_algorithm}\par\vspace{2pt}
\noindent
\hyperref[sec:appendix_prompts]{\textbf{G\hspace{0.8em}Prompt Templates}}\dotfill\pageref{sec:appendix_prompts}\par\vspace{2pt}
\noindent
\hyperref[sec:appendix_limitation]{\textbf{H\hspace{0.8em}Limitations and Discussion}}\dotfill\pageref{sec:appendix_limitation}

\endgroup

\section{Additional Analyses and Ablations}
\label{sec:appendix_ablation}

\subsection{Rank Fusion Strategy}
\label{sec:rank_fuse}

The main paper adopts Time-Weighted Reciprocal Rank Fusion (TW-RRF) to aggregate candidate lists across channels and turns. Here we compare it against five alternatives on WebVid-CoVR-Test to justify this choice:
(1)~{Static-RRF} fuses only the current turn's two channel lists using reciprocal rank with fixed channel weights, discarding all historical evidence;
(2)~{Uniform-RRF} aggregates across a sliding window like TW-RRF but assigns equal weight to every turn;
(3)~{Penalty-RRF} replaces the additive indicator with a penalty rank $r_{\mathrm{pen}}$ for absent candidates, so that missing from a list incurs a small negative contribution rather than zero;
(4)~{SimMax} performs score-level fusion by taking the maximum raw similarity across channels and turns;
(5)~{SimSum} sums the raw similarity scores instead.

As shown in Table~\ref{tab:rff_ab}, TW-RRF achieves the best R@1 from Turn~2 onward and the strongest final performance on every metric. Static-RRF is marginally ahead at Turn~1 but plateaus quickly because it cannot accumulate cross-turn evidence. Uniform-RRF performs close to TW-RRF in early turns, with slightly lower BRI at Turns~2--3, but falls behind as the interaction progresses, indicating that recency weighting matters more once later turns carry more refined intent. Penalty-RRF underperforms both Uniform-RRF and TW-RRF, indicating that penalizing absent candidates is overly conservative and suppresses useful partial evidence from individual channels. The two score-level baselines (SimMax, SimSum) are consistently weaker, which is expected because raw similarity scales differ across channels and turns, making direct score comparison unreliable. These results validate the three design choices behind TW-RRF: multi-turn aggregation, recency weighting, and rank-level (rather than score-level) fusion.

\begin{table*}[htbp]
    \caption{\textbf{Ablation on rank fusion strategies} on WebVid-CoVR-Test. We compare TW-RRF against three rank-level variants (Static-RRF, Uniform-RRF, Penalty-RRF) and two score-level baselines (SimMax, SimSum). TW-RRF has the best R@1 from Turn~2 onward and the best Turn-5 result on every metric; Static-RRF is marginally ahead in R@1 at Turn~1, and Uniform-RRF has slightly lower BRI at Turns~2--3.}
    \label{tab:rff_ab}
    \begin{center}
    \resizebox{0.999\linewidth}{!}{
    \begin{tabular}{l|ccccc|ccccc|ccccc}
\toprule[1pt]
\multicolumn{1}{c|}{\multirow{2}{*}{\begin{tabular}[c]{@{}c@{}}Rank Fusion\\ Strategy\end{tabular}}} & \multicolumn{5}{c|}{\textbf{R@1$\uparrow$}}                                   & \multicolumn{5}{c|}{\textbf{R@5$\uparrow$}}                                   & \multicolumn{5}{c}{\textbf{BRI$\downarrow$}}                                    \\
\multicolumn{1}{c|}{}                                                                         & \textit{Turn 1} & \textit{2} & \textit{3} & \textit{4} & \textit{5} & \textit{Turn 1} & \textit{2} & \textit{3} & \textit{4} & \textit{5} & \textit{Turn 1} & \textit{2} & \textit{3} & \textit{4} & \textit{5} \\ \hline
Penalty-RRF                                                                                   & 70.11           & 75.77      & 80.13      & 84.59      & 87.79      & 86.66           & 88.54      & 89.63      & 90.88      & 92.94      & 0.6944          & 0.5614     & 0.4885     & 0.4321     & 0.3877     \\
Static-RRF                                                                                    & 74.36           & 78.33      & 80.95      & 83.76      & 86.03      & 87.25           & 86.93      & 87.56      & 88.81      & 90.10      & 0.6701          & 0.5557     & 0.4915     & 0.4490     & 0.4164     \\
Uniform-RRF                                                                                   & 73.63           & 79.81      & 83.57      & 86.79      & 89.03      & 88.38           & 88.71      & 90.69      & 91.98      & 93.19      & 0.6726          & 0.5250     & 0.4512     & 0.4106     & 0.3736     \\
SimMax                                                                                        & 68.58           & 69.76      & 75.63      & 80.16      & 82.75      & 85.52           & 84.47      & 85.84      & 87.28      & 88.58      & 0.7162          & 0.6052     & 0.5520     & 0.5092     & 0.4738     \\
SimSum                                                                                        & 69.64           & 73.32      & 75.35      & 76.96      & 83.29      & 87.56           & 89.98      & 90.65      & 91.28      & 92.61      & 0.6938          & 0.5583     & 0.4933     & 0.4526     & 0.4181     \\
TW-RRF                                                                                        & 74.30           & 80.32      & 84.39      & 87.72      & 90.22      & 87.28           & 89.28      & 91.28      & 93.04      & 94.13      & 0.6737          & 0.5271     & 0.4538     & 0.4034     & 0.3643     \\ \bottomrule[1pt]
\end{tabular}
    }
\end{center}
\end{table*}

\begin{table*}[tbp]
    \caption{\textbf{Effect of transplanting Rank-cap into open-loop interactive baselines} on WebVid-CoVR-Test. Turn~0 is omitted because all methods share the same initialization. While Rank-cap consistently improves both \texttt{IVR} and \texttt{UMIVR}, ReCoVR remains the best across all turns, indicating that rank-space intervention alone does not account for the full gains.}
    \label{tab:baseline_rankcap}
    \centering
    \resizebox{0.7\linewidth}{!}{
        \begin{tabular}{ll|ccccc}
\toprule[1pt]
\multicolumn{2}{c|}{Methods} & \textit{Turn 1} & \textit{2} & \textit{3} & \textit{4} & \textit{5} \\ \hline
\multirow{5}{*}{\rotatebox{90}{\textbf{R@1$\uparrow$}}}
& IVR & 66.55 & 70.11 & 71.95 & 73.00 & 73.51 \\
& IVR + Rank-cap & 73.04 & 76.64 & 79.89 & 80.75 & 82.08 \\
& UMIVR & 67.10 & 71.09 & 72.85 & 74.10 & 75.04 \\
& UMIVR + Rank-cap & 72.81 & 77.03 & 80.63 & 81.77 & 83.18 \\
\rowcolor{tbgreen!70}
& ReCoVR (ours) & \textbf{74.30} & \textbf{80.32} & \textbf{84.39} & \textbf{87.72} & \textbf{90.22} \\ \hline
\multirow{5}{*}{\rotatebox{90}{\textbf{BRI$\downarrow$}}}
& IVR & 0.7368 & 0.6187 & 0.5560 & 0.5161 & 0.4880 \\
& IVR + Rank-cap & 0.7074 & 0.5715 & 0.4985 & 0.4516 & 0.4187 \\
& UMIVR & 0.7285 & 0.6049 & 0.5394 & 0.4986 & 0.4700 \\
& UMIVR + Rank-cap & 0.6967 & 0.5555 & 0.4798 & 0.4314 & 0.3982 \\
\rowcolor{tbgreen!70}
& ReCoVR (ours) & \textbf{0.6737} & \textbf{0.5271} & \textbf{0.4538} & \textbf{0.4034} & \textbf{0.3643} \\ \bottomrule[1pt]
\end{tabular}
    }
\end{table*}

\subsection{Baseline Rank-cap Transfer}
\label{sec:rankcap}

To isolate the contribution of the Rank-cap policy, we transplant the same rule into the open-loop interactive baselines \texttt{IVR} and \texttt{UMIVR}. Their original single-channel query-update pipelines remain unchanged. We reuse the same cap positions as ReCoVR ($\kappa_{\mathrm{neg}}{=}11$, $\kappa_{\mathrm{neu}}{=}4$) and derive the satisfaction signal from the simulator's action type, which gives the baselines a rejection signal that ReCoVR must instead infer from the feedback, applying demotion as a lightweight post-processing step without introducing the Intent Pathway or State Reflection.

As shown in Table~\ref{tab:baseline_rankcap}, Rank-cap yields consistent gains for both baselines across all turns: at Turn~5, R@1 improves by 8.57 points for \texttt{IVR} and 8.14 points for \texttt{UMIVR}, with corresponding BRI improvements. This confirms that rank-space intervention is a useful correction mechanism that transfers readily to existing interactive pipelines. ReCoVR achieves the best performance at every turn, and the margin grows over successive rounds: the gap over UMIVR + Rank-cap is 1.49 points at Turn~1 and 7.04 points at Turn~5. This widening trend is consistent with the component ablation in Table~\ref{tab:ablation}: Rank-cap provides effective immediate intervention, while the dual-channel Intent Pathway and trajectory-aware State Reflection contribute cumulative gains that become increasingly important as interaction history grows. The full architecture combines these complementary mechanisms to sustain improvement across turns.

\subsection{Statistical Significance via Paired Bootstrap}
\label{sec:paired_ci}

The main results in Table~\ref{tab:compare_covr} are point estimates. To attach uncertainty to the comparison against the strongest adapted baseline, we rerun ReCoVR and UMIVR in a single matched environment and pair all 2,556 test-session trajectories. Both methods start from the identical Turn-0 ranking, so every session yields a paired difference. We then resample complete session trajectories 10,000 times. Each resample uses the same session indices for both methods. We report percentile 95\% intervals on the per-turn difference in R@1.

Table~\ref{tab:paired_ci} reports the outcome. The paired improvement grows from $+6.77$ points at Turn~1 to $+14.32$ points at Turn~5, and every interval lies entirely above zero. This matched rerun reproduces the ordering and turn-wise trend of Table~\ref{tab:compare_covr} to within $0.55$ points for both methods, so we retain the original numbers as the main result and use the rerun only to quantify uncertainty. Two scope notes apply. The intervals characterize test-session sampling variability conditional on a fixed retriever, reasoning model, simulator, and interaction protocol; they do not capture variability from re-drawing model or simulator behavior. And the five turns are successive states of the same sessions rather than independent samples, so we report per-turn intervals instead of a single aggregate test.

\begin{table}[htbp]
    \caption{\textbf{Paired session-level uncertainty} for ReCoVR over UMIVR on WebVid-CoVR-Test, from a matched rerun of both methods over 2,556 paired session trajectories with 10,000 bootstrap resamples. The improvement widens across turns and every 95\% interval lies entirely above zero.}
    \label{tab:paired_ci}
    \begin{center}
    \resizebox{0.92\linewidth}{!}{
        \begin{tabular}{lccccc}
        \toprule[1pt]
                                  & \textit{Turn 1} & \textit{Turn 2} & \textit{Turn 3} & \textit{Turn 4} & \textit{Turn 5} \\ \hline
        UMIVR \textbf{R@1} $\uparrow$          & 67.21           & 71.56           & 73.40           & 74.61           & 75.35           \\
        \cellcolor{tbgreen!70}ReCoVR \textbf{R@1} $\uparrow$ & \cellcolor{tbgreen!70}73.98 & \cellcolor{tbgreen!70}79.93 & \cellcolor{tbgreen!70}85.05 & \cellcolor{tbgreen!70}88.18 & \cellcolor{tbgreen!70}89.67 \\ \hline
        $\Delta$\textbf{R@1} (pp)             & \textbf{+6.77}  & \textbf{+8.37}  & \textbf{+11.66} & \textbf{+13.58} & \textbf{+14.32} \\
        Paired 95\% CI            & [+5.20, +8.29]  & [+6.77, +9.90]  & [+10.09, +13.22] & [+12.01, +15.14] & [+12.75, +15.88] \\ \bottomrule[1pt]
        \end{tabular}
    }
    \end{center}
\end{table}

\subsection{Measuring Open-Loop Failure Modes}
\label{sec:failure_diag}

Our motivation rests on an empirical premise: open-loop interactive retrieval drifts, stagnates, and keeps returning candidates the user has already rejected. We measure these behaviors directly rather than inferring them from end-task metrics. We use the matched rerun of Appendix~\ref{sec:paired_ci} to score every \textit{active} feedback transition. A transition is active when the simulated user is still interacting and has not issued a stop action. We compare the target rank before and after each active turn. Turn~0, preparation steps, and carry-forward rows after an early stop are excluded. Because ReCoVR terminates more sessions earlier, the two methods enter interaction on the same 1,167 sessions but execute a different number of active transitions (3,010 versus 4,061), so all rates are computed against each method's own denominator.

Table~\ref{tab:failure_diag} reports the outcome, with each failure mode operationalized as a measurable proxy on the target rank. We define query drift as a turn in which the target rank strictly worsens. It occurs in 18.52\% of UMIVR's active transitions versus 12.69\% of ReCoVR's. Stagnation means that feedback fails to improve the target rank. Its rate is 54.25\% for UMIVR versus 20.23\% for ReCoVR. A rejected candidate is repeated when the top-1 result remains unchanged after non-stop feedback. This occurs in 70.77\% of UMIVR's transitions. ReCoVR has no rate to report for this failure mode because rank-cap demotes the rejected top-1 at every such turn and rules it out by design.

The more informative question is what happens \textit{after} a bad turn. Restricting to cases where the session continues and another feedback turn is executed, ReCoVR improves the target rank on the following turn in 88.15\% of cases against 30.12\% for UMIVR. This contrast should be read against each method's own base rate: ReCoVR improves the rank in 79.77\% of its turns overall and UMIVR in 45.75\%, so recovering at 88.15\% puts ReCoVR $8.38$ points \textit{above} its own average, while 30.12\% puts UMIVR $15.63$ points \textit{below} its own. This reversal in sign distinguishes a closed loop from an open one more clearly than the raw ratio: a stalled open-loop system tends to stay stalled, because nothing in its update rule registers that the previous turn failed. Finally, the rank-cap policy fires in 49.20\% of ReCoVR's active transitions, and 76.57\% of those transitions show a same-turn improvement in target rank. The same turn also performs intent decomposition, retrieval, and fusion, so this is an association attached to the triggering condition rather than an isolated causal effect of rank-cap; the matched ablations of Sec.~\ref{sec:ablation} carry the component-level attribution.

\begin{table}[htbp]
    \caption{\textbf{Direct measurement of the open-loop failure modes} over active feedback transitions on WebVid-CoVR-Test. Rates use each method's own number of active transitions (ReCoVR 3,010, UMIVR 4,061). Each failure mode from Sec.~\ref{sec:intro} is operationalized as a measurable proxy. Repeated rejected candidates cannot occur under ReCoVR, whose rank-cap rule demotes a rejected top-1 after every non-stop feedback. The lower block conditions on transitions that failed to improve the rank and are followed by another executed turn.}
    \label{tab:failure_diag}
    \begin{center}
    \resizebox{0.99\linewidth}{!}{
        \begin{tabular}{lcc}
        \toprule[1pt]
        Measured over active feedback transitions                                   & UMIVR              & \cellcolor{tbgreen!70}ReCoVR \\ \hline
        Query drift: the target rank strictly worsens $\downarrow$                  & 18.52\% (752/4,061)   & \cellcolor{tbgreen!70}\textbf{12.69\% (382/3,010)} \\
        Stagnation: the target rank does not improve $\downarrow$                   & 54.25\% (2,203/4,061) & \cellcolor{tbgreen!70}\textbf{20.23\% (609/3,010)} \\
        Repeated rejected candidates: top-1 unchanged after feedback $\downarrow$    & 70.77\%               & \cellcolor{tbgreen!70}\textit{prevented by rank-cap} \\ \hline
        \rowcolor{tbgrey!70}\multicolumn{3}{c}{\textit{After a turn that failed to improve the rank}} \\ \hline
        Target rank improves on the next turn $\uparrow$                            & 30.12\% (1,720 eligible) & \cellcolor{tbgreen!70}\textbf{88.15\% (557 eligible)} \\
        \quad relative to the method's own per-turn improvement rate                & $-15.63$ pp           & \cellcolor{tbgreen!70}\textbf{$+8.38$ pp} \\
        Median change in target rank on the next turn $\downarrow$                  & $0$                   & \cellcolor{tbgreen!70}\textbf{$-2$} \\ \bottomrule[1pt]
        \end{tabular}
    }
    \end{center}
\end{table}

\subsection{Robustness to Prompt Rewording and Structured-Output Reliability}
\label{sec:prompt_robust}

Intent decomposition, query reformulation, state reflection, and question asking are all realized as structured prompts, so we test whether the results depend on a particular wording. We create two additional semantically equivalent prompt suites called Variant~A and Variant~B. Each rewrites all four module prompts. We rerun the full protocol on WebVid-CoVR-Test with the retriever, simulator, decoding settings, and interaction protocol held fixed.

Table~\ref{tab:prompt_robust} reports the three trajectories. All suites preserve monotonic multi-turn improvement, and Turn-5 R@1 spans only $0.39$ points ($89.95$--$90.34$). The effect of rewording is concentrated in the early turns: Variant~B loses $1.60$ points at Turn~1 and then recovers to within $0.27$ points by Turn~5. We therefore claim robustness of the final conclusion under semantically equivalent wording, not insensitivity to prompt design in general.

We separately quantify how often the structured-output interface itself fails. Across the three suites, 22,408 schema-constrained generations are issued, of which 22,304 are valid on the first pass ($99.54\%$); the remaining 104 are recovered by a deterministic fallback, and no run terminates early or produces an unusable session. The Intent Decomposer and State Reflector drive the interaction loop. They produce valid outputs on the first pass in 18,903 of 18,904 calls ($99.995\%$). Reliability is not uniform across modules: under Variant~B the T2V initializer falls to $91.18\%$ first-pass validity, which is precisely the case the fallback exists to absorb. These statistics characterize output \textit{interface} reliability by measuring whether a decision can be parsed. They do not measure the semantic correctness of each decision. End-task metrics assess the resulting retrieval performance.

\begin{table}[htbp]
    \caption{\textbf{Robustness to prompt rewording} on WebVid-CoVR-Test. Variants~A and~B reword all four reasoning-module prompts while preserving their meaning. R@1 is reported per turn; the Turn-5 spread across the three suites is $0.39$ points.}
    \label{tab:prompt_robust}
    \begin{center}
    \resizebox{0.85\linewidth}{!}{
        \begin{tabular}{l|ccccc|c}
        \toprule[1pt]
        Prompt suite & \textit{Turn 1} & \textit{2} & \textit{3} & \textit{4} & \textit{5} & \textit{Turn-5} BRI$\downarrow$ \\ \hline
        \cellcolor{tbgreen!70}Original (main paper) & \cellcolor{tbgreen!70}74.30 & \cellcolor{tbgreen!70}80.32 & \cellcolor{tbgreen!70}84.39 & \cellcolor{tbgreen!70}87.72 & \cellcolor{tbgreen!70}90.22 & \cellcolor{tbgreen!70}0.3643 \\
        Variant A    & 74.26 & 79.73 & 84.12 & 87.37 & 90.34 & 0.3697 \\
        Variant B    & 72.70 & 78.76 & 83.02 & 86.90 & 89.95 & 0.3846 \\ \bottomrule[1pt]
        \end{tabular}
    }
    \end{center}
\end{table}

\subsection{Fixed-Reference versus Dynamic Anchoring}
\label{sec:anchor}

The CoVR channel anchors each turn on the previously fused top-1 result $\hat{L}_{t-1}(1)$ rather than on the original reference video $v_{\mathrm{ref}}$. The motivation is that a \texttt{Modify} instruction describes how the \textit{currently displayed} result should change, so applying it to the original reference would attach the edit to the wrong referent. We verify this choice with a matched comparison in which only the anchor differs: the simulator, reasoning VLM, retriever, both pathways, rank-cap, TW-RRF, turn budget, and stopping rule are unchanged, and each setting receives feedback generated online against its own displayed result rather than a replay of the other's dialogue.

Table~\ref{tab:anchor} shows that dynamic anchoring is better at all five interactive turns, by $+1.52$ to $+4.10$ points of R@1 and $+2.99$ on average. The advantage generally grows in later turns. It peaks at Turn~4 and narrows slightly at Turn~5. Both settings improve monotonically over turns, so a fixed reference does not break interactive refinement; it simply converges lower. We do not report an oracle-anchor variant, since selecting the anchor from target-side information is not available at retrieval time and would serve only as an upper bound.

\begin{table}[htbp]
    \caption{\textbf{Anchor choice for the CoVR channel} on WebVid-CoVR-Test (R@1). Only the visual anchor differs between the two settings; each receives online feedback on its own displayed result.}
    \label{tab:anchor}
    \begin{center}
    \resizebox{0.8\linewidth}{!}{
        \begin{tabular}{l|ccccc}
        \toprule[1pt]
        Anchor for the CoVR channel & \textit{Turn 1} & \textit{2} & \textit{3} & \textit{4} & \textit{5} \\ \hline
        Fixed reference $v_{\mathrm{ref}}$ & 72.78 & 77.93 & 81.25 & 83.62 & 86.41 \\
        \cellcolor{tbgreen!70}Dynamic: previous fused top-1 (ours) & \cellcolor{tbgreen!70}\textbf{74.30} & \cellcolor{tbgreen!70}\textbf{80.32} & \cellcolor{tbgreen!70}\textbf{84.39} & \cellcolor{tbgreen!70}\textbf{87.72} & \cellcolor{tbgreen!70}\textbf{90.22} \\ \hline
        $\Delta$ & +1.52 & +2.39 & +3.14 & +4.10 & +3.81 \\ \bottomrule[1pt]
        \end{tabular}
    }
    \end{center}
\end{table}

\subsection{Simplified Multi-Turn Strategies}
\label{sec:fullhistory}

Sec.~\ref{sec:ablation} compares ReCoVR against a baseline that rewrites the complete interaction history into a single composed query at every turn. Table~\ref{tab:attribution} places that baseline alongside the other simplified strategies studied in this appendix, each of which keeps one mechanism of the closed loop and discards the rest. There is little history to exploit at Turn~1, and all variants are competitive at this stage. They fall behind ReCoVR as interaction proceeds. Current-turn-only fusion is marginally ahead at Turn~1 but falls $4.19$ points behind by Turn~5. Table~\ref{tab:fullhistory} then gives the full trajectory of the history-rewriting baseline.

\begin{table}[htbp]
    \caption{\textbf{Simplified multi-turn strategies} on WebVid-CoVR-Test. Each row retains one closed-loop mechanism and discards the rest. The retained mechanisms are history aggregation, rank-space correction, and single-turn fusion.}
    \label{tab:attribution}
    \begin{center}
    \resizebox{0.88\linewidth}{!}{
        \begin{tabular}{lccc}
        \toprule[1pt]
        Strategy                                                                            & \textit{Turn 1} & \textit{Turn 5} & Gap at Turn 5 \\ \hline
        Full-history query rewriting (CoVR-only, no Reflection)                             & 65.54 & 77.43 & $-12.79$ \\
        Rank-cap transplanted onto UMIVR (Appendix~\ref{sec:rankcap})                       & 72.81 & 83.18 & $-7.04$ \\
        ReCoVR without Rank-cap (Sec.~\ref{sec:ablation})                                   & 70.42 & 83.26 & $-6.96$ \\
        Current-turn-only rank fusion, no cross-turn evidence (Appendix~\ref{sec:rank_fuse}) & 74.36 & 86.03 & $-4.19$ \\ \hline
        \rowcolor{tbgreen!70}\textbf{ReCoVR}                                                & \textbf{74.30} & \textbf{90.22} & --- \\ \bottomrule[1pt]
        \end{tabular}
    }
    \end{center}
\end{table}
 The baseline uses the same retriever, reasoning VLM, simulator, and interaction protocol, retrieves through the CoVR channel only, and has no Reflection Pathway; it differs from the rolling CoVR-only variant of Table~\ref{tab:ablation}, which updates the previous rewritten query incrementally rather than regenerating it from the whole history. Its R@1 improves across turns, so history does carry usable signal, but its gains at deeper cutoffs are small: R@5 stays below ReCoVR's Turn-1 value at every turn, R@10 stays nearly flat, R@50 falls below its Turn-0 value, and BRI improves only slowly. Re-deriving a single query from an ever-longer history tends to over-condition the ranking on whichever constraints the model chooses to surface, which is the failure that trajectory-level diagnosis and rank-level fusion are designed to avoid.

\begin{table}[htbp]
    \caption{\textbf{Full-history rewriting baseline} on WebVid-CoVR-Test: same setup as ReCoVR, CoVR channel only, no Reflection Pathway, with the query regenerated from the complete interaction history at every turn. Deltas are against ReCoVR under the matched run.}
    \label{tab:fullhistory}
    \begin{center}
    \resizebox{0.85\linewidth}{!}{
        \begin{tabular}{l|cccccc}
        \toprule[1pt]
        Metric & \textit{Turn 0} & \textit{1} & \textit{2} & \textit{3} & \textit{4} & \textit{5} \\ \hline
        R@1 $\uparrow$   & 54.30 & 65.54 & 70.58 & 73.24 & 75.04 & 77.43 \\
        R@5 $\uparrow$   & 77.00 & 76.56 & 78.25 & 78.96 & 80.95 & 82.94 \\
        R@10 $\uparrow$  & 84.35 & 83.22 & 84.15 & 83.10 & 84.94 & 85.56 \\
        R@50 $\uparrow$  & 95.77 & 95.19 & 94.13 & 94.25 & 93.63 & 93.04 \\
        BRI $\downarrow$ & --    & 0.7816 & 0.6806 & 0.6160 & 0.5689 & 0.5309 \\ \hline
        $\Delta$R@1 vs.\ ReCoVR & 0.00 & $-8.76$ & $-9.74$ & $-11.15$ & $-12.68$ & $-12.79$ \\ \bottomrule[1pt]
        \end{tabular}
    }
    \end{center}
\end{table}

\section{Generalization}
\label{sec:generalization}

Because ReCoVR is training-free and operates above frozen retrievers, its usefulness depends on whether the gains survive changes of domain, retrieval backbone, task, and reasoning model; we evaluate each in turn.

\textbf{Generalization across domains.}
Table~\ref{tab:fashioniq} evaluates cross-domain transfer on FashionIQ~\citep{fashioniq}, a composed image retrieval benchmark with three fashion categories. Although ReCoVR starts from a weaker Turn~0 retriever than recent single-turn CIR methods, interactive feedback steadily improves performance and surpasses them after several turns. Compared with the adapted interactive baseline UMIVR, ReCoVR achieves larger and more sustained gains across all categories, while UMIVR tends to plateau after early turns. This suggests that closed-loop correction is not tied to video-specific cues, but can also support fine-grained composed image retrieval.

\begin{table*}[htbp]
    \caption{\textbf{Generalization to composed image retrieval} on FashionIQ. ReCoVR consistently outperforms single-turn or interactive methods across all categories.}
    \label{tab:fashioniq}

    \begin{center}
    \resizebox{0.99\linewidth}{!}{
    \begin{tabular}{cccccccccccccc}
\toprule[1pt]
\multicolumn{2}{c|}{\multirow{2}{*}{Methods}}                         & \multicolumn{4}{c|}{{Dress}}                 & \multicolumn{4}{c|}{{Shirts}}               & \multicolumn{4}{c}{{Tops\&Tees}} \\
\multicolumn{2}{c|}{}                                                & \textbf{R@1$\uparrow$}   & \textbf{R@10$\uparrow$}  & \textbf{R@50$\uparrow$}  & \multicolumn{1}{c|}{\textbf{BRI$\downarrow$}}    & \textbf{R@1$\uparrow$}   & \textbf{R@10$\uparrow$}  & \textbf{R@50$\uparrow$}  & \multicolumn{1}{c|}{\textbf{BRI$\downarrow$}}   & \textbf{R@1$\uparrow$}      & \textbf{R@10$\uparrow$}     & \textbf{R@50$\uparrow$}    & \textbf{BRI$\downarrow$}     \\ \hline
\rowcolor{tbgrey!70}\multicolumn{14}{c}{\textit{Single-Turn Composed Image Retrieval}}                                                                                                                                                        \\ \hline
\multicolumn{2}{l|}{CoVR-2~\citep{webvid_covr2}$_{TPAMI~24}$}                                          & -     & 46.53 & 69.60 & \multicolumn{1}{c|}{-}      & -     & 51.23 & 70.64 & \multicolumn{1}{c|}{-}     & -        & 52.14    & 73.27   & -       \\
\multicolumn{2}{l|}{iSHARLE~\citep{agnolucci2025isearle}$_{TPAMI~25}$}                                         & -     & 31.81 & 50.20 & \multicolumn{1}{c|}{-}      & -     & 24.19 & 45.12 & \multicolumn{1}{c|}{-}     & -        & 31.72    & 53.29   & -       \\
\multicolumn{2}{l|}{X$^R$~\citep{yang2026xr}$_{WWW~26}$}                                              & -     & 38.91 & 56.82 & \multicolumn{1}{c|}{-}      & -     & 28.71 & 52.50 & \multicolumn{1}{c|}{-}     & -        & 43.91    & 62.57   & -       \\
\multicolumn{2}{l|}{REFINE~\citep{refine}$_{TOMM~26}$}                                          & -     & 50.32 & 72.98 & \multicolumn{1}{c|}{-}      & -     & 57.90 & 77.87 & \multicolumn{1}{c|}{-}     & -        & 59.82    & 80.83   & -       \\ \hline
\rowcolor{tbgreen!70}\multicolumn{14}{c}{\textit{Interactive Refinement (BLIP-ViT-Large)}}                                                                                                                                                                      \\ \hline
\multirow{6}{*}{\rotatebox{90}{UMIVR}}        & \multicolumn{1}{c|}{\textit{Turn 0}} & 8.73  & 26.18 & 45.46 & \multicolumn{1}{c|}{-}      & 9.03  & 30.32 & 47.25 & \multicolumn{1}{c|}{-}     & 11.22    & 31.36    & 52.83   & -       \\
                              & \multicolumn{1}{c|}{\textit{Turn 1}} & 15.32 & 38.47 & 60.59 & \multicolumn{1}{c|}{3.535}  & 16.39 & 41.51 & 60.94 & \multicolumn{1}{c|}{3.530} & 18.05    & 44.01    & 64.71   & 3.290   \\
                              & \multicolumn{1}{c|}{\textit{Turn 2}} & 18.59 & 40.80 & 62.92 & \multicolumn{1}{c|}{3.228}  & 18.55 & 42.89 & 62.81 & \multicolumn{1}{c|}{3.229} & 20.40    & 45.49    & 66.96   & 2.998   \\
                              & \multicolumn{1}{c|}{\textit{Turn 3}} & 20.92 & 41.45 & 63.46 & \multicolumn{1}{c|}{3.045}  & 20.07 & 42.93 & 62.90 & \multicolumn{1}{c|}{3.063} & 22.44    & 45.84    & 67.77   & 2.831   \\
                              & \multicolumn{1}{c|}{\textit{Turn 4}} & 22.76 & 40.70 & 62.92 & \multicolumn{1}{c|}{2.926}  & 20.76 & 42.69 & 63.35 & \multicolumn{1}{c|}{2.954} & 23.41    & 46.46    & 68.28   & 2.717   \\
                              & \multicolumn{1}{c|}{\textit{Turn 5}} & 24.00 & 42.64 & 64.55 & \multicolumn{1}{c|}{2.830} & 21.59 & 43.47 & 63.84 & \multicolumn{1}{c|}{2.871} & 24.48    & 47.83    & 68.79   & 2.631   \\ \hline
\multirow{6}{*}{\rotatebox{90}{ReCoVR(Ours)}} & \multicolumn{1}{c|}{\textit{Turn 0}} & 8.73  & 26.18 & 45.46 & \multicolumn{1}{c|}{-}      & 9.03  & 30.32 & 47.25 & \multicolumn{1}{c|}{-}     & 11.22    & 31.36    & 52.83   & -       \\
                              & \multicolumn{1}{c|}{\textit{Turn 1}} & 23.30 & 49.23 & 71.54 & \multicolumn{1}{c|}{3.257}  & 26.74 & 49.56 & 71.15 & \multicolumn{1}{c|}{3.268} & 26.77    & 53.54    & 74.55   & 3.035   \\
                              & \multicolumn{1}{c|}{\textit{Turn 2}} & 29.25 & 51.81 & 73.83 & \multicolumn{1}{c|}{2.815}  & 31.50 & 51.13 & 72.77 & \multicolumn{1}{c|}{2.834} & 33.86    & 55.12    & 76.64   & 2.616   \\
                              & \multicolumn{1}{c|}{\textit{Turn 3}} & 35.75 & 56.52 & 77.14 & \multicolumn{1}{c|}{2.580}  & 37.54 & 56.82 & 76.79 & \multicolumn{1}{c|}{2.603} & 40.13    & 60.17    & 80.01   & 2.393   \\
                              & \multicolumn{1}{c|}{\textit{Turn 4}} & 41.99 & 60.14 & 79.67 & \multicolumn{1}{c|}{2.407}  & 43.72 & 59.72 & 79.20 & \multicolumn{1}{c|}{2.434} & 46.20    & 63.13    & 81.95   & 2.230   \\
                              & \multicolumn{1}{c|}{\textit{Turn 5}} & \textbf{47.89} & \textbf{62.62} & \textbf{81.41} & \multicolumn{1}{c|}{\textbf{2.268}}  & \textbf{47.60} & \textbf{61.92} & \textbf{80.57} & \multicolumn{1}{c|}{\textbf{2.301}} & \textbf{50.99}    & \textbf{66.34}    & \textbf{82.66}   & \textbf{2.101}   \\ \bottomrule[1pt]
\end{tabular}

    }
\end{center}
\end{table*}

\textbf{Generalization across retrieval backbones.}\label{sec:retr_backbone}
We examine whether ReCoVR depends on a particular retrieval encoder. Table~\ref{tab:diff_channel} evaluates nine CoVR/T2V channel configurations on WebVid-CoVR-Test, including seven homogeneous settings where both channels use the same encoder and two heterogeneous settings where the channels use different encoders. These configurations cover CLIP~\citep{radford2021learning}, BLIP~\citep{li2022blip}, LanguageBind~\citep{zhu2023languagebind}, and Qwen3-VL-Embedding~\citep{qwen3vlembedding} families.

All configurations show consistent improvement over turns, indicating that ReCoVR's gains are not tied to a specific retrieval backbone. Stronger encoders provide better starting points and higher final accuracy, while weaker encoders also benefit substantially from feedback, suggesting that closed-loop correction can compensate for limited initial retrieval capacity. Importantly, even when the Turn~0 retriever is already competitive with recent single-turn CoVR methods, ReCoVR still provides substantial additional gains, showing that \textbf{interactive refinement is complementary to advances in single-turn retrieval}. The heterogeneous settings further show that the dual-channel design does not require matched encoders and can also operate with encoder diversity.

\begin{table*}[htbp]
    \caption{\textbf{ReCoVR with different retrieval backbones} on WebVid-CoVR-Test. We evaluate seven homogeneous configurations (same encoder for both channels) spanning five model families, plus two heterogeneous configurations (different encoders across channels). All configurations show consistent multi-turn improvement, confirming that ReCoVR is retrieval-backbone-agnostic.}
    \label{tab:diff_channel}
    \begin{center}
    \resizebox{1.0\linewidth}{!}{
    \begin{tabular}{ll|cccccc|cccccc|ccccc}
\toprule[1pt]
\multirow{2}{*}{\begin{tabular}[c]{@{}l@{}}CoVR\\ Channel\end{tabular}} & \multirow{2}{*}{\begin{tabular}[c]{@{}l@{}}T2V\\ Channel\end{tabular}} & \multicolumn{6}{c|}{\textbf{R@1$\uparrow$}}                                                & \multicolumn{6}{c|}{\textbf{R@5$\uparrow$}}                                                & \multicolumn{5}{c}{\textbf{BRI$\downarrow$}}                                    \\
                                                                        &                                                                        & \textit{Turn 0} & \textit{1} & \textit{2} & \textit{3} & \textit{4} & \textit{5} & \textit{Turn 0} & \textit{1} & \textit{2} & \textit{3} & \textit{4} & \textit{5} & \textit{Turn 1} & \textit{2} & \textit{3} & \textit{4} & \textit{5} \\ \hline
CLIP-Base                                                               & CLIP-Base                                                              & 46.32           & 65.30      & 72.77      & 77.31      & 80.87      & 83.88      & 68.47           & 81.30      & 84.12      & 86.50      & 88.07      & 89.12      & 0.9293          & 0.7433     & 0.6791     & 0.5859     & 0.5379     \\
CLIP-Large                                                              & CLIP-Large                                                             & 55.36           & 73.67      & 80.91      & 84.90      & 88.18      & 90.02      & 77.78           & 87.40      & 90.06      & 91.90      & 93.35      & 94.21      & 0.6328          & 0.4961     & 0.4246     & 0.3761     & 0.3394     \\
BLIP-Base                                                               & BLIP-Base                                                              & 51.17           & 70.62      & 77.07      & 81.96      & 85.52      & 88.26      & 74.02           & 85.76      & 87.36      & 89.83      & 91.63      & 92.57      & 0.7598          & 0.6051     & 0.5263     & 0.4717     & 0.4291     \\
BLIP-Large                                                              & BLIP-Large                                                             & 54.30           & 74.30      & 80.32      & 84.39      & 87.72      & 90.22      & 77.00           & 87.28      & 89.28      & 91.28      & 93.04      & 94.13      & 0.6737          & 0.5271     & 0.4538     & 0.4034     & 0.3643     \\
LanguageBind                                                            & LanguageBind                                                           & 43.11           & 71.36      & 77.35      & 81.18      & 83.41      & 85.99      & 66.28           & 87.95      & 87.90      & 88.93      & 90.30      & 91.08      & 0.8897          & 0.6584     & 0.5585     & 0.4962     & 0.4504     \\
Qwen3VL-Embed2B                                                         & Qwen3VL-Embed2B                                                        & 59.47           & 79.97      & 84.00      & 88.34      & 91.20      & 92.72      & 81.30           & 92.14      & 93.35      & 94.64      & 95.74      & 96.60      & 0.5194          & 0.3965     & 0.3353     & 0.2931     & 0.2615     \\
Qwen3VL-Embed8B                                                         & Qwen3VL-Embed8B                                                        & 64.36           & 82.47      & 86.93      & 90.02      & 92.80      & 94.33      & 85.02           & 93.86      & 94.09      & 95.74      & 96.64      & 97.38      & 0.4262          & 0.3225     & 0.2711     & 0.2358     & 0.2091     \\
BLIP-Large                                                              & LanguageBind                                                           & 54.30           & 76.41      & 82.55      & 87.36      & 90.30      & 92.33      & 77.00           & 91.08      & 91.86      & 93.27      & 94.68      & 95.81      & 0.6408          & 0.4780     & 0.3988     & 0.3461     & 0.3071     \\
Qwen3VL-Embed2B                                                         & BLIP-Base                                                              & 59.47           & 75.74      & 82.75      & 86.62      & 89.32      & 91.74      & 81.30           & 89.16      & 91.71      & 93.43      & 94.76      & 95.70      & 0.5469          & 0.4317     & 0.3691     & 0.3256     & 0.2919     \\ \bottomrule[1pt]
\end{tabular}
    }
\end{center}
\end{table*}

\begin{table*}[htbp]
\begin{minipage}[b]{0.48\textwidth}
    \caption{\textbf{Generalization to I-TVR} on MSRVTT-1kA and AVSD-1k-Test. Adding the Reflection Pathway to UMIVR improves retrieval and outperforms all baselines on both video benchmarks.}
    \label{tab:compare_t2v}
    \centering
    \resizebox{\linewidth}{!}{
        \begin{tabular}{clcccccc}
\toprule[1pt]
\multicolumn{2}{l}{Methods}                                 & \textit{Turn 0} & \textit{1}     & \textit{2}     & \textit{3}     & \textit{4}     & \textit{5}     \\ \hline
\rowcolor{tbgrey!70}\multicolumn{8}{c}{\textit{MSRVTT-1kA Dataset}}                                                                                                                    \\ \hline
\multirow{4}{*}{\rotatebox{90}{\textbf{R@1$\uparrow$}}} & \multicolumn{1}{l|}{IVR}    & 43.1            & 61.4           & 65.8           & 68.2           & 69.3           & 69.9           \\
                              & \multicolumn{1}{l|}{Merlin} & 43.1            & 50.3           & 54.9           & 56.3           & 57.9           & 59.6           \\
                              & \multicolumn{1}{l|}{UMIVR}  & 43.1            & 65.0           & 74.3           & 81.3           & 84.9           & 87.3           \\
                    \rowcolor{tbgreen!70}\cellcolor{white}          & \multicolumn{1}{l|}{Ours}   & 43.1            & \textbf{70.2}  & \textbf{83.7}  & \textbf{89.8}  & \textbf{92.4}  & \textbf{93.5}  \\ \hline
\multirow{4}{*}{\rotatebox{90}{\textbf{BRI$\downarrow$}}} & \multicolumn{1}{l|}{IVR}    & -               & 1.050          & 0.876          & 0.793          & 0.742          & 0.707          \\
                              & \multicolumn{1}{l|}{Merlin} & -               & 1.230          & 1.127          & 1.054          & 0.999          & 0.953          \\
                              & \multicolumn{1}{l|}{UMIVR}  & -               & 0.985          & 0.751          & 0.614          & 0.519          & 0.451          \\
                \rowcolor{tbgreen!70}\cellcolor{white}              & \multicolumn{1}{l|}{Ours}   & -               & \textbf{0.946} & \textbf{0.675} & \textbf{0.524} & \textbf{0.429} & \textbf{0.365} \\ \hline
\rowcolor{tbgrey!70}\multicolumn{8}{c}{\textit{AVSD-1k-Test Dataset}}                                                                                                                  \\ \hline
\multirow{4}{*}{\rotatebox{90}{\textbf{R@1$\uparrow$}}} & \multicolumn{1}{l|}{IVR}    & 30.3            & 36.0           & 38.8           & 40.6           & 41.3           & 41.5           \\
                              & \multicolumn{1}{l|}{Merlin} & 30.3            & 34.9           & 37.5           & 38.9           & 40.3           & 41.3           \\
                              & \multicolumn{1}{l|}{UMIVR}  & 30.3            & 43.2           & 50.4           & 56.0           & 59.9           & 63.2           \\
                \rowcolor{tbgreen!70}\cellcolor{white}              & \multicolumn{1}{l|}{Ours}   & 30.3            & \textbf{47.6}  & \textbf{64.2}  & \textbf{71.3}  & \textbf{74.8}  & \textbf{76.6}  \\ \hline
\multirow{4}{*}{\rotatebox{90}{\textbf{BRI$\downarrow$}}} & \multicolumn{1}{l|}{IVR}    & -               & 1.802          & 1.698          & 1.638          & 1.599          & 1.570          \\
                              & \multicolumn{1}{l|}{Merlin} & -               & 1.838          & 1.743          & 1.670          & 1.613          & 1.566          \\
                              & \multicolumn{1}{l|}{UMIVR}  & -               & 1.665          & 1.438          & 1.272          & 1.146          & 1.050          \\
                \rowcolor{tbgreen!70}\cellcolor{white}              & \multicolumn{1}{l|}{Ours}   & -               & \textbf{1.580} & \textbf{1.265} & \textbf{1.049} & \textbf{0.912} & \textbf{0.819} \\ \bottomrule[1pt]
\end{tabular}
    }
\par\vspace{0pt}
\end{minipage}
\hfill
\begin{minipage}[b]{0.495\textwidth}
    \caption{\textbf{Effect of backbone VLMs} on WebVid-CoVR-Test. ReCoVR shows consistent multi-turn improvement across two model families (Qwen3-VL and InternVL3.5) and two model sizes (4B and 8B). These results support the framework's backbone-agnostic design.}
    \label{tab:basevlm}
    \centering
    \resizebox{\linewidth}{!}{
        \begin{tabular}{ll|ccccc}
\toprule[1pt]
\multicolumn{2}{c|}{$\bullet~$\textbf{BaseVLMs}}           & \textit{Turn 1} & \textit{2} & \textit{3} & \textit{4} & \textit{5} \\ \hline
\multirow{4}{*}{\rotatebox{90}{\textbf{R@1$\uparrow$}}}  & Qwen3-VL-4B    & 74.30           & 80.32      & 84.39      & 87.72      & 90.22      \\
                      & Qwen3-VL-8B    & 74.84           & 81.69      & 85.80      & 89.12      & 91.88      \\
                      & InternVL3.5-4B & 75.59           & 82.94      & 87.6       & 90.61      & 92.64      \\
                      & InternVL3.5-8B & 77.74           & 84.31      & 88.30      & 91.00      & 93.00      \\ \hline
\multirow{4}{*}{\rotatebox{90}{\textbf{R@5$\uparrow$}}}  & Qwen3-VL-4B    & 87.28           & 89.28      & 91.28      & 93.04      & 94.13      \\
                      & Qwen3-VL-8B    & 89.48           & 90.45      & 92.10      & 93.43      & 94.60      \\
                      & InternVL3.5-4B & 87.91           & 91.71      & 93.86      & 94.99      & 96.09      \\
                      & InternVL3.5-8B & 88.15           & 92.06      & 94.09      & 95.42      & 96.21      \\ \hline
\multirow{4}{*}{\rotatebox{90}{\textbf{R@10$\uparrow$}}} & Qwen3-VL-4B    & 93.08           & 93.81      & 94.56      & 95.27      & 96.24      \\
                      & Qwen3-VL-8B    & 93.43           & 94.01      & 95.11      & 96.40      & 96.91      \\
                      & InternVL3.5-4B & 91.55           & 95.19      & 96.32      & 96.87      & 97.57      \\
                      & InternVL3.5-8B & 91.74           & 95.15      & 96.40      & 97.26      & 97.77      \\ \hline
\multirow{4}{*}{\rotatebox{90}{\textbf{BRI$\downarrow$}}}  & Qwen3-VL-4B    & 0.6737          & 0.5271     & 0.4538     & 0.4034     & 0.3643     \\
                      & Qwen3-VL-8B    & 0.6483          & 0.4884     & 0.4111     & 0.3601     & 0.3224     \\
                      & InternVL3.5-4B & 0.6365          & 0.4653     & 0.3803     & 0.3260     & 0.2868     \\
                      & InternVL3.5-8B & 0.6271          & 0.4543     & 0.3708     & 0.3172     & 0.2784     \\ \bottomrule[1pt]
\end{tabular}
    }
\par\vspace{0pt}
\end{minipage}
\end{table*}

\textbf{Generalization to interactive text-to-video retrieval.}\label{sec:itvr_transfer}
We evaluate whether the closed-loop Reflection Pathway transfers beyond ICoVR to standard interactive text-to-video retrieval (I-TVR). To isolate this effect, we integrate the Reflection Pathway into UMIVR as a plug-in module, while keeping the original I-TVR setting and feedback protocol. The interaction terminates once the target is found; otherwise, the simulated user follows the VideoQA-based feedback protocol used in prior I-TVR methods~\citep{ivr_auto,merlin,umivr}.

As shown in Table~\ref{tab:compare_t2v}, adding the Reflection Pathway consistently improves the open-loop UMIVR pipeline on both MSRVTT-1kA and AVSD-1k-Test. The gains are especially large on AVSD-1k-Test, where the lower Turn~0 accuracy leaves more room for trajectory-level correction. This suggests that the closed-loop diagnosis mechanism is not specific to composed retrieval, but can also strengthen existing interactive pipelines.

\textbf{Generalization across reasoning VLMs.}\label{sec:reasoning_vlm}
We evaluate whether ReCoVR depends on a particular reasoning VLM. In this experiment, the user simulator is fixed to Qwen3-VL-4B, while ReCoVR uses different reasoning VLMs from two families and two scales: Qwen3-VL-4B/8B and InternVL3.5-4B/8B. This setting separates the simulator from the retrieval agent and tests whether the observed gains require using the same VLM family as the simulator.

Table~\ref{tab:basevlm} shows that all reasoning VLMs produce steady multi-turn improvements. Larger models generally achieve higher final performance, but switching to a different model family also preserves the gain pattern. This suggests that ReCoVR's improvement comes from the closed-loop reasoning structure rather than simulator-agent model sharing. The results further indicate that the framework is robust to the choice and scale of the reasoning VLM. Qwen3-VL-4B is one of the smallest models in this comparison and is the default for every main-paper result. The reported gains therefore do not require the larger 8B variants.

\section{Efficiency Analysis}
\label{sec:appendix_efficiency}

The closed loop comes at a computational price: every ReCoVR turn runs intent decomposition and trajectory reflection on the VLM, adding about $1.6$s per executed turn over the strongest adapted baseline. Faster convergence partially offsets this overhead, but overall system time remains higher. Two properties keep this cost manageable. First, the overhead lies in reasoning rather than retrieval: nearest-neighbor search and rank fusion together take hundredths of a second per turn, so the cost scales with the number of VLM calls rather than with gallery size, and shrinks with a smaller or faster reasoning model. Second, reading results and writing feedback dominate each turn in the real-user latency estimate. Under a conservative estimate of human response time, the entire ReCoVR system accounts for roughly a seventh of end-to-end latency. The following analysis reports the per-turn breakdown, memory footprint, and deployment estimate.

All efficiency experiments in this section are conducted on a \textbf{single NVIDIA RTX A6000 GPU}. We evaluate on {100 instances} sampled from WebVid-CoVR-Test (the same set used in the Sec.~\ref{sec:user_study}) under the same setup as the main experiments (up to five interactive turns). Among them, {57 instances} fail at Turn~0 and thus enter the interactive stage. The interaction terminates early once the target video is ranked first, so fewer executed turns directly indicate faster convergence. Unless otherwise specified, all reported timings are measured from the interactive pipeline used in automatic evaluation. For ReCoVR, the full first-turn query reform cost is included in the reported Turn~1 latency, matching the definition used throughout the paper.

\begin{table}[hbp]
\centering
\caption{\textbf{Method-level interactive cost on 100 instances.} We report total interactive time, the number of executed turns, the average number of executed turns per active query, and the average end-to-end time per executed turn.}
\label{tab:appendix_eff_e2e}
\resizebox{\linewidth}{!}{
\begin{tabular}{l|c|c|c|c}
\toprule
Method & Total Time (s, 100q) & Executed Turns & Avg Turns / Active Query & Avg Time / Executed Turn (s) \\
\midrule
IVR & 482.06 & 223 & 3.91 & 2.16 \\
UMIVR & 637.15 & 231 & 4.05 & 2.76 \\
\textbf{ReCoVR (ours)} & {908.04} & {208} & {3.65} & {4.37} \\
\bottomrule
\end{tabular}}
\vspace{-1em}
\end{table}

Table~\ref{tab:appendix_eff_e2e} provides a view of end-to-end efficiency beyond a simple average with 100 instances, since the number of executed turns varies under early stopping. ReCoVR has the largest total runtime, but executes the {fewest} interactive turns overall: 208 turns, compared with more than 220 turns for both adapted baselines. Thus, its larger total runtime is not due to longer interaction trajectories, but to higher per-turn computation. This is expected because each ReCoVR turn performs dual-channel retrieval and closed-loop reasoning, whereas the baselines use lighter open-loop query updates. Overall, ReCoVR trades higher per-turn cost for faster convergence in terms of executed turns.

\begin{table}[t]
\centering
\caption{\textbf{Per-turn latency breakdown of ReCoVR.} Time is averaged over all executed interactive turns ($N{=}208$). Percentages are computed against the measured end-to-end turn cost in the automatic pipeline. Individual module percentages may not sum exactly to the subtotal due to rounding.}
\label{tab:appendix_eff_modules}
\resizebox{\linewidth}{!}{
\begin{tabular}{l|c|c}
\toprule
Module & Avg Time / Turn (s) & Ratio in Measured Pipeline (\%) \\
\midrule
Question Generation & 0.158 & 3.62 \\
Intent Decomposer & 0.950 & 21.76 \\
Query Reformer & 0.619 & 14.17 \\
T2V Retrieval & 0.017 & 0.39 \\
CoVR Retrieval & 0.011 & 0.25 \\
TW-RRF & 0.001 & 0.02 \\
Reflection & 0.847 & 19.41 \\
Simulated User Feedback & 1.767 & 40.48 \\
\midrule
System Total (ReCoVR only) & 2.598 & 59.52 \\
Measured Total / Turn & 4.365 & 100.00 \\
Estimated Total / Turn with Real Human Feedback (15s) & 17.598 & - \\
\bottomrule
\end{tabular}}
\end{table}

\begin{table}[t]
\centering
\caption{\textbf{Memory-bank footprint and offline bank-construction latency.} We report Resident Set Size (RSS) after model loading, after bank loading, and at peak during execution, along with the bank size in RAM and on disk, and the mean cost of constructing one bank entry on a cache miss.}
\label{tab:appendix_eff_memory}
\resizebox{\linewidth}{!}{
\begin{tabular}{l|ccc|cc|c}
\toprule
Method & RSS After Model (MB) & RSS After Bank (MB) & RSS Peak (MB) & Bank RAM (MB) & Bank Disk (MB) & Build Mean (s) \\
\midrule
IVR & 1857.95 & 1940.98 & 2281.07 & 0.98 & 0.54 & 0.908 \\
UMIVR & 1857.91 & 1943.28 & 2274.54 & 0.98 & 0.54 & 0.908 \\
\textbf{ReCoVR (ours)} & 1857.81 & 1944.11 & 2157.97 & 3.50 & 2.01 & 1.658 \\
\bottomrule
\end{tabular}}
\end{table}

Table~\ref{tab:appendix_eff_modules} explains where the per-turn latency of ReCoVR comes from. In the automatic pipeline, the largest single component is the simulated user, which takes {1.77s} per turn and accounts for {40.48\%} of the total latency. On the model side, the main cost comes from language-driven reasoning modules rather than retrieval itself: \textit{Intent Decomposer} takes 0.95s, \textit{Reflection} takes 0.85s, and \textit{Query Reformer} takes 0.62s per turn, whereas the combined cost of T2V retrieval, CoVR retrieval, and TW-RRF is only about 0.03s. This confirms that the computational overhead of ReCoVR is dominated by its reasoning and diagnosis stages, not by nearest-neighbor retrieval.

More importantly, Table~\ref{tab:appendix_eff_modules} shows why raw simulator timing should not be over-interpreted for realistic deployments. If the simulated user is replaced by a conservative estimate of {15 seconds} of real human feedback per turn, the end-to-end turn cost becomes {17.60s}. Under this setting, the user contributes about {85.24\%} of the latency budget, while the entire ReCoVR system contributes only {14.76\%}. ReCoVR is computationally heavier than the baselines. Human response time nevertheless dominates practical interaction time under this estimate.

Table~\ref{tab:appendix_eff_memory} shows that the memory bank adds only a small footprint relative to model residency. For all methods, the bank primarily stores the long-term Progress Memory $\mathit{PM}^L$, which caches pre-computed video captions for the gallery. ReCoVR's bank is slightly larger ({3.50MB} in RAM, {2.01MB} on disk) because $\mathit{PM}^L$ additionally stores richer metadata used by the Reflection Pathway, compared with the caption-only banks of IVR and UMIVR ({0.98MB}). In both cases, the bank is negligible relative to the ${\sim}$1.86GB base model footprint, occupying less than {0.19\%} of model memory. The short-term Progress Memory $\mathit{PM}^S$ and the Result Memory RM are session-level structures that are created and discarded per query, contributing no persistent storage cost.

The table also reports the offline cost of bank construction on cache misses. ReCoVR requires {1.658s} on average to build one $\mathit{PM}^L$ entry, while IVR and UMIVR require {0.908s}. Importantly, these are {offline amortized} costs: once built, the cached entries are reused across all sessions and queries, so they do not affect the online per-turn latency reported in Tables~\ref{tab:appendix_eff_e2e} and~\ref{tab:appendix_eff_modules}.

Overall, ReCoVR is not the fastest method in raw system time, which is the expected cost of dual-channel retrieval and closed-loop reasoning. However, it converges in fewer turns, its retrieval operations are extremely lightweight, and its bank footprint is negligible. Most importantly, in realistic use the dominant latency comes from human feedback rather than model computation, making ReCoVR's efficiency trade-off well justified by its retrieval gains.

\section{User Simulation and Human-Feedback Study Details}
\label{sec:supp_user_sim}

\subsection{Motivation: Why Simulate Users?}
\label{sec:supp_sim_motivation}

Evaluating interactive retrieval systems faces a well-known tension between reproducibility and realism~\citep{zhang2024usimagent, dou2025simulatorarena}. Offline evaluation on static benchmarks (e.g., measuring Recall or NDCG per turn) assumes that all preceding turns have been perfectly resolved, ignoring the cascading effects of errors across turns~\citep{dalton2020trec}. Online user studies capture realistic interaction dynamics but are expensive, difficult to scale, and hard to reproduce. This gap has motivated the development of \textit{user simulators} that can interact with retrieval systems in a controlled, automated loop~\citep{zhang2024usimagent,azzopardi2007building, degachi2025towards, aliannejadi2024interactions}.

In the information retrieval community, user simulation has evolved from early rule-based or probabilistic models that estimate click and query reformulation behaviors~\citep{mo2025survey} from search logs~\citep{azzopardi2007building}, to recent LLM-based approaches that leverage instruction following and reasoning capabilities to role-play as users with specific information needs~\citep{zhang2024usimagent,wang2024depth,erbacher2022interactive,balog2023user}. For instance, USimAgent~\citep{zhang2024usimagent} employs an LLM to generate queries, predict clicks, and decide when to stop searching, demonstrating that LLM-based simulators can produce coherent multi-round search sessions that approximate real user behavior. These advances have made simulated-user evaluation a standard paradigm for interactive systems where large-scale human studies are impractical.

In interactive text-to-video retrieval (I-TVR), existing methods adopt a question-answering interaction protocol: the system generates a clarifying question at each turn, and a user simulator provides an answer based on the ground-truth target~\citep{ivr_auto,merlin,umivr}. This protocol has enabled reproducible multi-turn evaluation across I-TVR benchmarks.

However,~\textbf{composed video retrieval presents a fundamentally different challenge}. Unlike text-only retrieval where user intent is expressed through text alone, ICoVR requires the user to coordinate feedback across two modalities, referencing both visual content and textual modifications simultaneously. The user must observe a video result, compare it against a mental model of the target, and articulate the remaining visual and semantic gap, which demands a vision-language understanding capability that text-only simulators cannot provide. Furthermore, user feedback in ICoVR is not limited to answering predefined questions; it may take the form of relative modifications (e.g., ``\texttt{make the scene brighter}''), direct descriptions (e.g., ``\texttt{I want a beach at sunset}''), or answers to system-generated questions, requiring a flexible simulation framework. No prior work has addressed user simulation for the composed video retrieval setting. These considerations motivate us to design a user simulator tailored for ICoVR, which we describe next.

\begin{figure*}[tbp]
    \centering
    \includegraphics[width=1.0\linewidth]{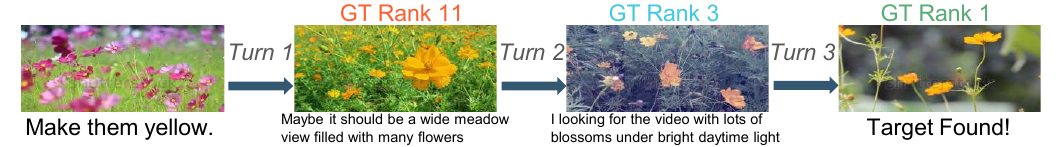}
    \vspace{-1em}
    \caption{\textbf{Qualitative trajectory with human feedback.}
A real user progressively refines the search intent across three turns, improving the ground-truth rank from 11 to 3 and finally to 1.}
\label{fig:user_case}
\vspace{-1em}
\end{figure*}

\subsection{User Simulator Design}
\label{sec:supp_sim_design}

Our simulator is built on a vision-language model that acts as a surrogate user throughout the interactive session. The overall procedure is summarized in Algorithm~\ref{alg:user_sim}. At the start of each evaluation query, the simulator receives the ground-truth target video $v_{\mathrm{gt}}$ and generates an internal caption to establish a persistent visual memory of the target (Line~\ref{line:caption}). The simulator caches this caption and the target video representation for reuse across all subsequent turns. This keeps its representation of the search goal consistent.

At each interaction turn $t$, the simulator receives the system's top-ranked candidate $\hat{L}_t(1)$ and optionally a clarifying question $q_t$. If the candidate matches the ground truth, the session terminates (Line~\ref{line:stop}). Otherwise, the simulator enters a two-stage process. In the \textit{analysis stage} (Line~\ref{line:analyze}), the VLM jointly observes both the target and the candidate video and produces a structured comparison, classifying the discrepancy as either \texttt{MINOR} (the candidate is semantically related but differs in specific attributes) or \texttt{MAJOR} (the candidate is largely irrelevant to the target). 
In the \textit{feedback generation stage} (Lines~\ref{line:action_start}--\ref{line:action_end}), the simulator selects an action type based on the analysis outcome and generates natural-language feedback $u_t$ accordingly. This follows recent user simulation protocols that separate target understanding from feedback generation~\citep{zhang2024usimagent}, while adapting the action space to the relative and descriptive feedback that composed retrieval invites. The action depends on the target, the displayed result, and whether a question was asked, not on the method being evaluated:
\begin{itemize}
    \item \textbf{Modify}: When the discrepancy is minor, the simulator suggests a targeted modification to the current result (e.g., ``\texttt{change the background to outdoor}'').
    \item \textbf{Rewrite}: When the discrepancy is major, the simulator generates a fresh description of the target video.
    \item \textbf{Answer}: When a system question $q_t$ is available and the discrepancy is major, the simulator may choose to answer the question based on its knowledge of the target.
\end{itemize}
This action selection produces a mix of relative and descriptive feedback, mirroring the diverse interaction patterns of real users. All methods in our experiments interact with the same simulator under identical prompting configurations, ensuring a fair comparison.

\begin{algorithm}[tbp]
\caption{User Simulator for ICoVR}
\label{alg:user_sim}
\begin{algorithmic}[1]
\REQUIRE Ground-truth target video $v_\mathrm{gt}$; VLM $\mathcal{V}$; max turns $T$
\STATE $\mathrm{caption} \leftarrow \mathcal{V}.\mathrm{describe}(v_\mathrm{gt})$ \hfill \COMMENT{Cache target representation} \label{line:caption}
\FOR{$t = 1$ \TO $T$}
    \STATE Receive candidate $\hat{L}_t(1)$ and optional question $q_t$ from system
    \IF{$\hat{L}_t(1) = v_\mathrm{gt}$} \label{line:stop}
        \STATE \textbf{return} \texttt{STOP}
    \ENDIF
    \STATE $(\mathrm{status},\;\mathrm{analysis}) \leftarrow \mathcal{V}.\mathrm{compare}(v_\mathrm{gt},\; \hat{L}_t(1))$ \label{line:analyze}
    \hfill \COMMENT{$\mathrm{status} \in \{\texttt{MINOR},\, \texttt{MAJOR}\}$}
    \IF{$\mathrm{status} = \texttt{MINOR}$} \label{line:action_start}
        \STATE $u_t \leftarrow \mathcal{V}.\mathrm{modify}(\mathrm{analysis},\; v_\mathrm{gt})$
    \ELSIF{$q_t \neq \varnothing$ \AND $\mathrm{rand}() < 0.5$}
        \STATE $u_t \leftarrow \mathcal{V}.\mathrm{answer}(q_t,\; v_\mathrm{gt})$
    \ELSE
        \STATE $u_t \leftarrow \mathcal{V}.\mathrm{rewrite}(v_\mathrm{gt})$ \label{line:action_end}
    \ENDIF
    \STATE Send feedback $u_t$ to the retrieval system
\ENDFOR
\end{algorithmic}
\end{algorithm}

\subsection{Human-Feedback Interface}
\label{sec:supp_user_interface}

\begin{figure*}[t]
    \centering
    \includegraphics[width=0.7\linewidth]{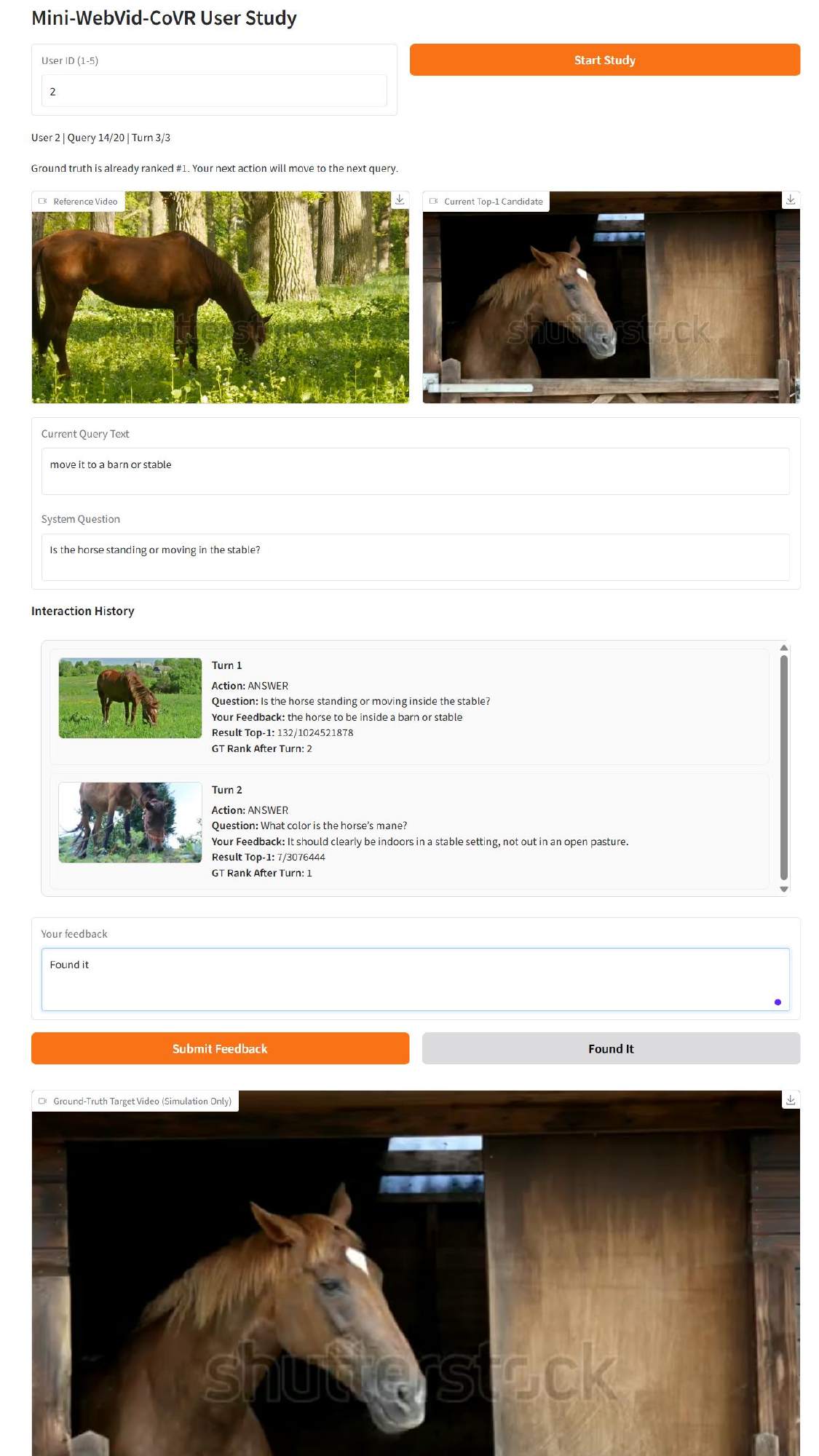}
    \caption{\textbf{Human-feedback interface.} Participants view the reference video, current top-ranked candidate, initial modification text, and an optional system question. They provide free-form feedback in the text box or click ``Found It'' to terminate the session. The interaction history panel tracks the search progression, and the ground-truth target video is shown at the bottom as the intended search goal.}
    \label{fig:user_interface}
\end{figure*}

To support the human-feedback evaluation of Sec.~\ref{sec:user_study} of the main paper, we develop a web-based interaction interface (Fig.~\ref{fig:user_interface}). The human-feedback study received institutional ethics approval. Participants were informed of the task procedure and minimal risks, and no demographic attributes or sensitive personal information were collected. The interface keeps the interaction setting close to the simulator-based evaluation protocol while allowing participants to provide free-form natural language feedback. At each turn, the participant views: (1)~the reference video from the original CoVR instance, (2)~the current top-ranked candidate returned by ReCoVR, (3)~the initial modification text, and (4)~an optional clarifying question generated by the system. A text box allows the participant to write natural-language feedback describing how the current result should be modified. The participant can also click a ``\texttt{Found It}'' button to indicate that the target has been located, which terminates the session early.

An interaction history panel records the previous turns, including the system question, the participant's feedback, and the resulting top-ranked candidate. This allows the participant to track how the search evolves over the session. For this controlled setting, the ground-truth target video is displayed in a separate section at the bottom as the intended search goal. Each participant is assigned a unique user ID and completes up to 20 instances, with a maximum of 3 interaction turns per instance. Fig.~\ref{fig:user_case} further illustrates a real interaction where the user progressively refines the search intent and guides ReCoVR to the target in three turns.

\subsection{Robustness to Degraded Simulated Feedback}
\label{sec:sim_robustness}

The main experiments use a target-conditioned simulator to provide scalable and repeatable feedback during offline evaluation. While this controlled protocol enables fair comparison across methods, real users may provide noisy, incomplete, sparse, or strategy-dependent feedback. We therefore stress-test ReCoVR by systematically degrading the simulated feedback on WebVid-CoVR-Test along three axes: \textit{surface noise}, which corrupts wording via character- and word-level perturbations~\citep{qiu2022benchmarking}; \textit{constraint dropping}, which randomly removes partial semantic constraints from the feedback; and \textit{behavioral policy shifts}, which replace the default mixed strategy with a single fixed interaction mode. These settings cover common non-ideal feedback conditions such as noisy wording, missing constraints, and sparse local edits.

\begin{table*}[tbp]
    \vspace{-1em}
    \caption{\textbf{Robustness to degraded simulated feedback} on WebVid-CoVR-Test. We report Turn~1 R@1, Turn~5 R@1, the interactive gain from Turn~0 to Turn~5, and BRI at Turn~5. Among the genuinely degraded settings (\textit{surface noise}, \textit{constraint dropping}, and \textit{modify-only}), ReCoVR remains consistently stronger than UMIVR and preserves large multi-turn gains. The \textit{rewrite-only} setting is included as a strategy-shift control.}
    \label{tab:sim_robustness}
    \centering
    \resizebox{0.98\linewidth}{!}{
    \begin{tabular}{ll|cccc|cccc}
    \toprule[1pt]
    \multirow{2}{*}{Category} & \multirow{2}{*}{Setting} & \multicolumn{4}{c|}{UMIVR} & \multicolumn{4}{c}{ReCoVR (ours)} \\
     &  & \textit{Turn 1} & \textit{Turn 5} & \textit{Gain@5} & \textit{BRI@5} & \textit{Turn 1} & \textit{Turn 5} & \textit{Gain@5} & \textit{BRI@5} \\ \hline
    Clean & Default simulator & 67.10 & 75.04 & 20.74 & 0.4700 & {74.30} & {90.22} & {35.92} & {0.3643} \\ \hline
    \multirow{2}{*}{Surface Noise}
      & Light: word swap + synonym replacement & 66.98 & 74.65 & 20.35 & 0.4745 & {73.90} & {89.01} & {34.71} & {0.3757} \\
      & Heavy: word deletion + character swap & 64.91 & 74.06 & 19.76 & 0.4978 & {71.40} & {88.73} & {34.43} & {0.3942} \\ \hline
    \multirow{2}{*}{Constraint Dropping}
      & Mild drop ($p{=}0.3$) & 66.71 & 74.77 & 20.47 & 0.4818 & {72.50} & {89.28} & {34.98} & {0.3781} \\
      & Severe drop ($p{=}0.5$) & 66.35 & 73.83 & 19.53 & 0.4876 & {72.30} & {89.08} & {34.78} & {0.3805} \\ \hline
    \multirow{2}{*}{Behavioral Policy Shift}
      & Modify-only & 60.02 & 64.05 & 9.75 & 0.6726 & {69.48} & {86.15} & {31.85} & {0.4405} \\
      & Rewrite-only (control) & 71.40 & 78.21 & 23.91 & 0.4023 & {76.72} & {92.14} & {37.84} & {0.2975} \\ \bottomrule[1pt]
    \end{tabular}
    \vspace{-1em}
    }
\end{table*}

Across all genuinely degraded conditions, ReCoVR's interactive gain remains within the narrow band of 31.85--34.98 points, compared with only 9.75--20.47 for UMIVR. Under surface-level perturbations and partial constraint omission, ReCoVR's Turn~5 R@1 stays within 1.49 points of its clean performance, with BRI degradation contained to at most 0.03. This tolerance is consistent with the design of the Reflection Pathway, which accumulates constraints across turns and corrects rankings from the retrieval trajectory rather than relying solely on the literal content of the latest feedback.

The most revealing test is the \textit{modify-only} policy, which forces the simulator to provide only local edits while suppressing any direct description of the target video. This severely limits the information available per turn. Under this condition, UMIVR's interactive gain collapses to just 9.75 points, whereas ReCoVR still achieves a gain of 31.85 points, reaching 86.15\% R@1 at Turn~5. This gap directly reflects the value of the dual-channel Intent Pathway: even when user feedback contains only relative modifications, the T2V channel can still construct effective standalone queries from the accumulated semantic context in Progress Memory, compensating for the absence of explicit target descriptions.

By contrast, the \textit{rewrite-only} policy improves both methods beyond their clean performance. Initial modification texts on WebVid-CoVR average only 4.6 words. A complete target description at each turn gives the retriever strictly more information than a partial edit. We include this condition as a strategy-shift control rather than a degradation case. The default simulator uses a mixed strategy to model the diversity of real user behavior rather than to maximize retrieval performance under any single mode.

Both behavioral policy conditions also bound how much ReCoVR depends on its optional Question Asking module. In these conditions the simulator follows a fixed interaction mode and never conditions on the generated question $q_t$, so they are equivalent to running the system with question asking disabled. ReCoVR reaches 86.15\% and 92.14\% R@1 at Turn~5 in the two cases, which indicates that the closed loop is driven by the retrieval trajectory rather than by whether the user chooses to answer.

These stress tests complement the human-feedback evaluation in Table~\ref{tab:user_study} of the main paper. Human participants provide free-form and less template-like feedback, while the degraded simulator tests systematically noisy, incomplete, and sparse feedback conditions. Together, they indicate that ReCoVR's advantage does not depend on perfectly informative feedback.

\section{Qualitative Case Studies}
\label{sec:supp_case_study}

\begin{figure*}[t]
    \centering
    \includegraphics[width=0.9\linewidth]{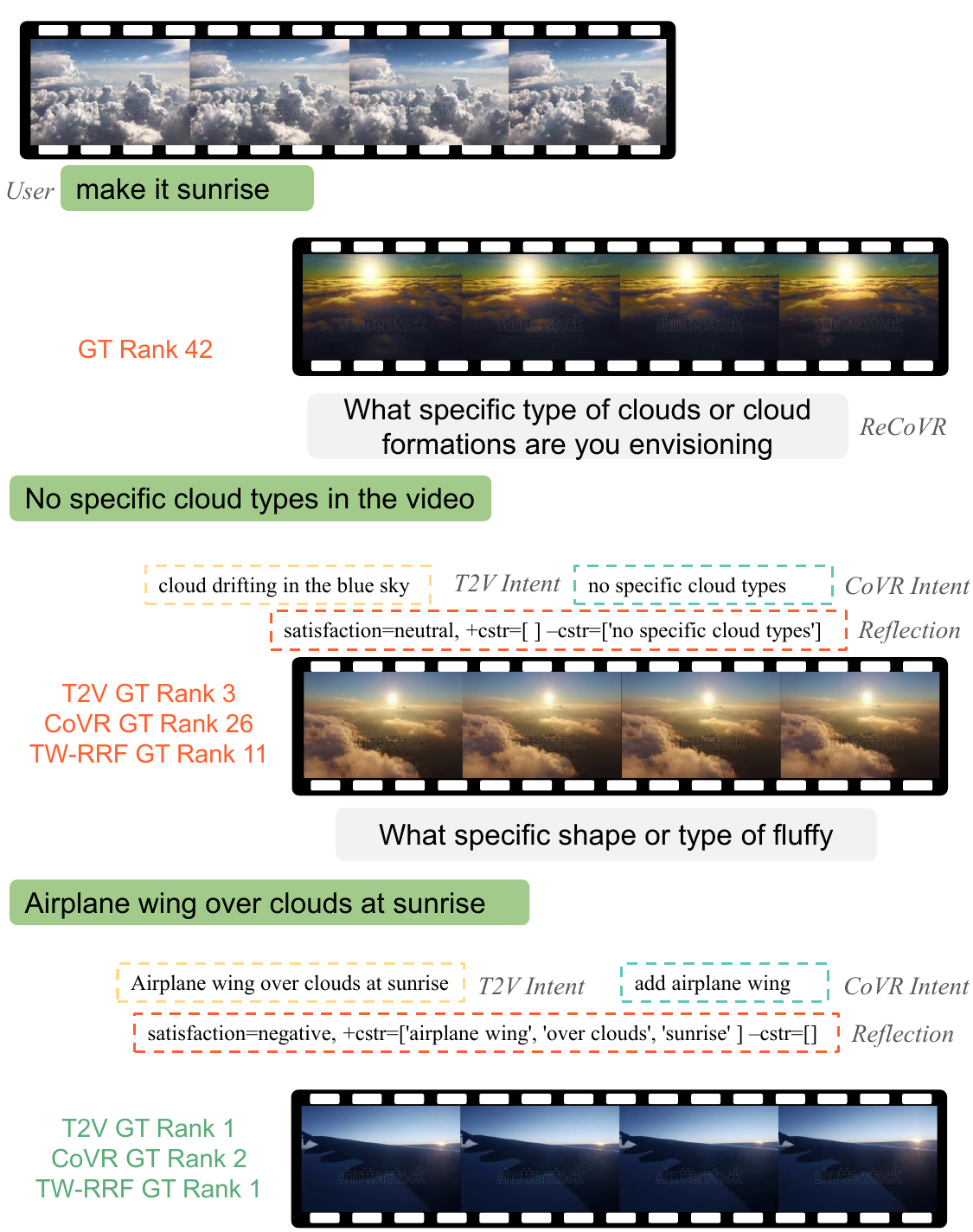}
    \caption{\textbf{Successful case on WebVid-CoVR.} The user progressively specifies scene details across two turns, improving the target from rank~42 to rank~1. The T2V channel drives convergence after the user provides a holistic scene description in Turn~2.}
    \label{fig:case_study_webvid}
\end{figure*}

\begin{figure*}[t]
    \centering
    \includegraphics[width=0.9\linewidth]{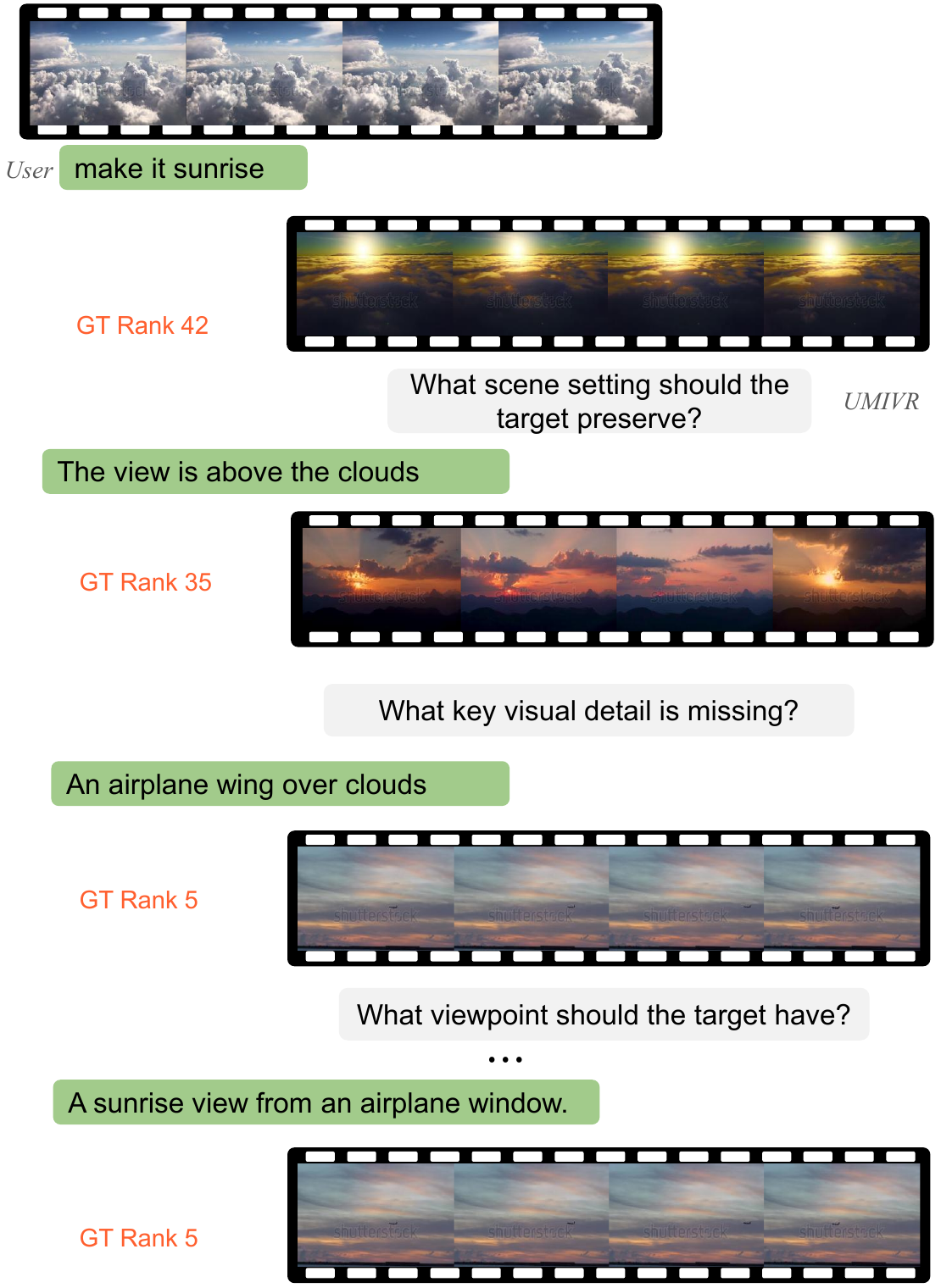}
    \caption{\textbf{Comparison on the same WebVid-CoVR case.} 
While ReCoVR improves the target from rank~42 to rank~1 in Fig.~\ref{fig:case_study_webvid}, UMIVR remains stuck at rank~5 under the same target instance and interaction budget, despite receiving explicit scene, airplane-wing, and viewpoint feedback. 
This comparison highlights the importance of trajectory-level reflection for correcting stalled retrieval.}
    \label{fig:baseline_case_study}
\end{figure*}

\begin{figure*}[t]
    \centering
    \includegraphics[width=0.9\linewidth]{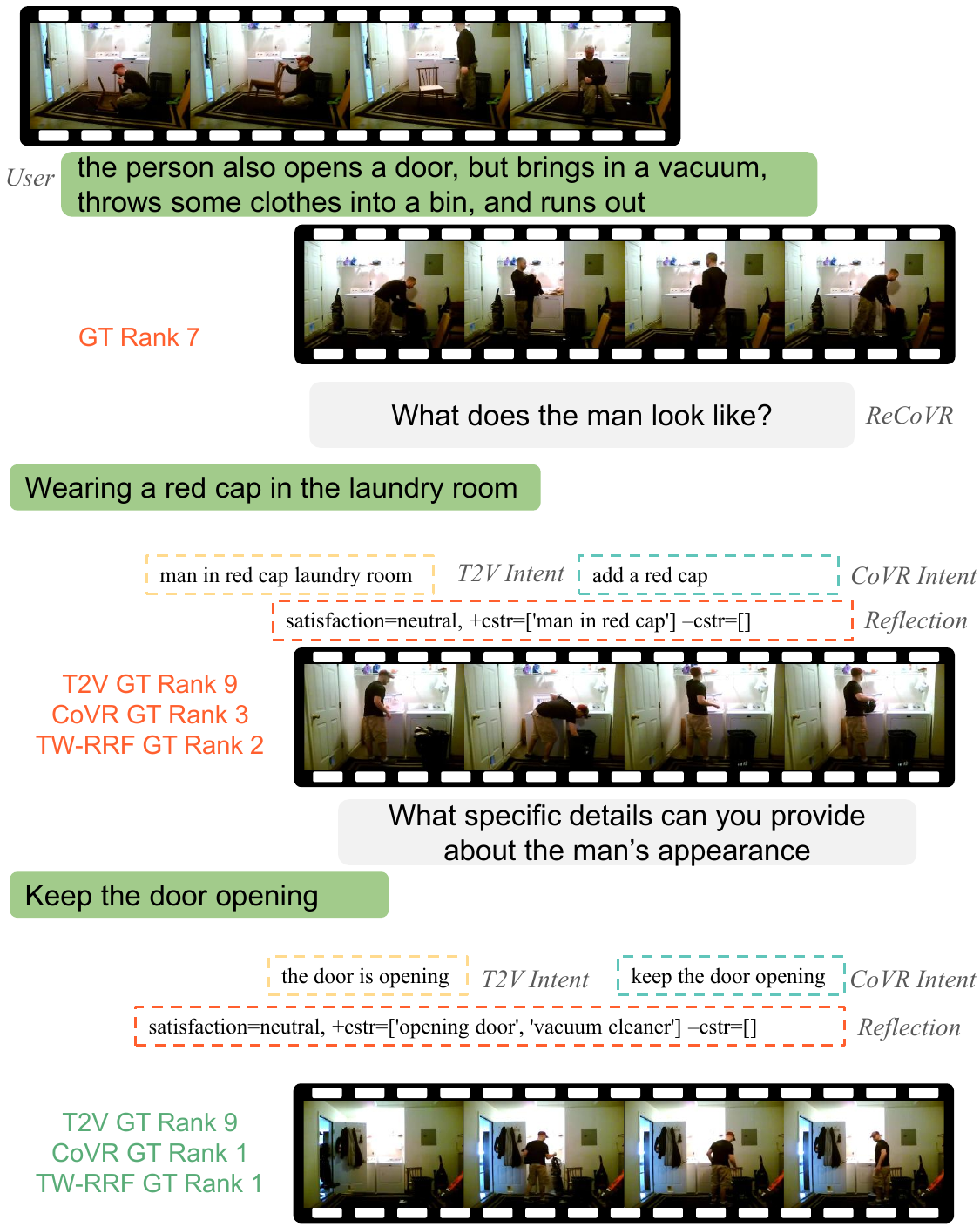}
    \caption{\textbf{Successful case on FineCVR.} Fine-grained appearance and action cues are anchored to the visual reference via the CoVR channel, improving the target from rank~7 to rank~1 over two interaction turns.}
    \label{fig:case_study_finecvr}
\end{figure*}

We present two successful examples and one same-instance baseline comparison to illustrate ReCoVR's behavior in practice. Fig.~\ref{fig:case_study_webvid} shows a WebVid-CoVR query where the user starts with ``\texttt{make it sunrise}'' (GT rank 42) and progressively specifies the scene across two turns. The Reflection Pathway first extracts a negative constraint to suppress irrelevant cloud types ($\sigma_1{=}\mathrm{neutral}$), then accumulates positive constraints [\textit{airplane wing},...] after an explicit rejection ($\sigma_2{=}\mathrm{negative}$), driving TW-RRF to rank~1. In contrast, Fig.~\ref{fig:baseline_case_study} shows the same target instance under UMIVR, which remains stuck at rank~5 despite explicit scene, airplane-wing, and viewpoint feedback. Fig.~\ref{fig:case_study_finecvr} illustrates a FineCVR case with a complex initial modification text (GT rank 7). Here the CoVR channel plays the dominant role: fine-grained appearance and action cues (``\texttt{red cap},'' ``\texttt{door opening}'') are best expressed relative to the visual reference, allowing the CoVR channel to reach rank~1 at Turn~2. Together, these cases demonstrate the complementarity of the dual channels and the role of reflection: T2V excels with holistic scene descriptions, CoVR is more effective for fine-grained reference-anchored refinements, and trajectory-level reflection helps correct stalled retrieval.

\begin{figure*}[t]
    \centering
    \includegraphics[width=0.9\linewidth]{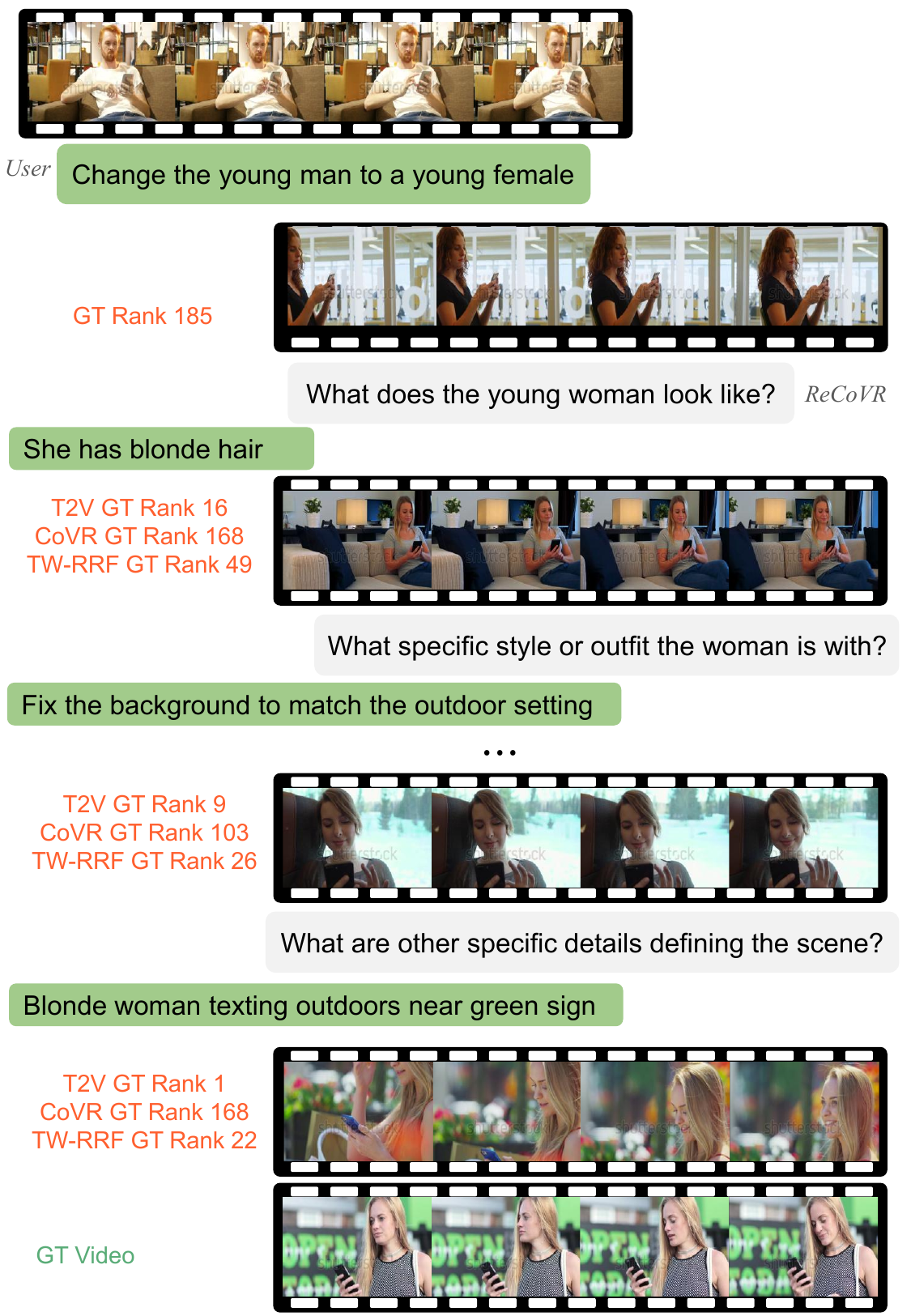}
    \caption{\textbf{Failure case.} A large visual gap between the reference (indoor, male) and target (outdoor, female) leaves the CoVR channel with a final target rank of~168. Despite the T2V channel reaching rank~1, fixed fusion yields a final rank of only~22.}
    \label{fig:failure_case}
\end{figure*}

Fig.~\ref{fig:failure_case} presents a case where the reference video (indoor, male subject) differs drastically from the target (outdoor, blonde female near a green sign). Despite multiple turns of user feedback, the CoVR channel fails to retrieve the target, ending at rank~168 because composed retrieval anchors to a visual reference that shares almost no similarity with the target. Although the T2V channel reaches rank~1 by the final turn, the fixed fusion strategy cannot suppress the CoVR channel's noisy contribution, yielding a final TW-RRF rank of only 22. This exposes a limitation of the current design: when the reference-target gap is large, adaptive channel weighting that down-weights the less reliable channel could improve fusion quality, which we leave for future work.

\section{Algorithmic Overview of ReCoVR}
\label{sec:appendix_algorithm}

The method description in the main paper is organized by pathway. Algorithm~\ref{alg:recovr} summarizes the end-to-end execution flow. It shows the data dependencies among modules and when each memory component is read and updated.

\renewcommand{\floatpagefraction}{0.85}
\begin{algorithm}[t!]
\caption{ReCoVR: Interactive Composed Video Retrieval}
\label{alg:recovr}
\begin{algorithmic}[1]
\REQUIRE Reference video $v_{\mathrm{ref}}$, modification text $u_0$, gallery $\mathcal{G}$, max turns $T$
\ENSURE Final ranked list $\hat{L}$
\STATE \COMMENT{\textit{Initialization}}
\STATE $\mathit{PM}^L \leftarrow \mathrm{CacheGalleryCaptions}(\mathcal{G})$
\STATE $\mathit{PM}^S \leftarrow \{v_{\mathrm{ref}},\; u_0,\; +\mathrm{cstr}{=}\varnothing,\; -\mathrm{cstr}{=}\varnothing\}$
\STATE $\mathit{RM} \leftarrow \varnothing$
\STATE \COMMENT{\textit{Turn 0: standard single-turn CoVR}}
\STATE $\hat{L}_0 \leftarrow \mathrm{CoVR}(v_{\mathrm{ref}},\, u_0,\, \mathcal{G})$;\quad $\mathit{RM} \leftarrow \mathit{RM} \cup \{(0,\, \hat{L}_0)\}$
\FOR{$t = 1$ \TO $T$}
    \STATE \COMMENT{\textit{User interaction}}
    \STATE $q_t \leftarrow \mathrm{AskQuestion}(\mathit{PM}^S)$
    \STATE Present $\hat{L}_{t-1}(1)$ and $q_t$ to the user; receive feedback $u_t$
    \IF{the user accepts $\hat{L}_{t-1}(1)$}
        \STATE \textbf{return} $\hat{L}_{t-1}$
    \ENDIF
    \STATE \COMMENT{\textit{Intent Pathway (uses $\mathit{PM}^S_{t-1}$ which includes $\Delta c_{t-1}$)}}
    \STATE $(\mathrm{search}_t,\, \mathrm{edit}_t) \leftarrow \mathrm{IntentDecomposer}(u_t,\, \mathit{PM}^S,\, \mathit{PM}^L)$
    \STATE $(\mathrm{search}_t,\, \mathrm{edit}_t) \leftarrow \mathrm{QueryReformer}(\mathrm{search}_t,\, \mathrm{edit}_t,\, \mathit{PM}^S)$
    \STATE \COMMENT{\textit{Reflection Pathway (independent of current Intent)}}
    \STATE $(\sigma_t,\, \Delta c_t) \leftarrow \mathrm{StateReflector}(u_t,\, \mathit{PM}^S,\, \mathit{PM}^L)$
    \STATE $\Delta r_t \leftarrow \mathrm{RankCapPolicy}(\sigma_t,\, \mathit{RM})$ \hfill \COMMENT{demote rejected/stagnant candidates}
    \STATE $\mathit{PM}^S \leftarrow \mathrm{UpdateConstraints}(\mathit{PM}^S,\, \Delta c_t)$
    \STATE \COMMENT{\textit{Dual-channel retrieval}}
    \STATE $L_t^{\mathrm{T2V}} \leftarrow \mathrm{T2V}(\mathrm{search}_t,\, \mathcal{G})$
    \STATE $L_t^{\mathrm{CoVR}} \leftarrow \mathrm{CoVR}\!\bigl(\hat{L}_{t-1}(1),\, \mathrm{edit}_t,\, \mathcal{G}\bigr)$ \hfill \COMMENT{dynamic anchoring}
    \STATE $(L_t^{\mathrm{T2V}},\, L_t^{\mathrm{CoVR}}) \leftarrow \mathrm{ApplyRankCap}(L_t^{\mathrm{T2V}},\, L_t^{\mathrm{CoVR}},\, \Delta r_t)$
    \STATE \COMMENT{\textit{Fusion and memory update}}
    \STATE $\mathit{RM} \leftarrow \mathit{RM} \cup \{(t,\, L_t^{\mathrm{T2V}},\, L_t^{\mathrm{CoVR}})\}$
    \STATE $\hat{L}_t \leftarrow \mathrm{TW\mbox{-}RRF}(\mathit{RM})$
    \STATE $\mathit{PM}^S \leftarrow \mathrm{UpdateProgress}(\mathit{PM}^S,\, \mathrm{search}_t,\, \mathrm{edit}_t,\, \hat{L}_t(1))$
\ENDFOR
\STATE \textbf{return} $\hat{L}_T$
\end{algorithmic}
\end{algorithm}

At each interactive turn, the Intent Pathway (Lines~14--16) and the Reflection Pathway (Lines~17--20) read the same snapshot $\mathit{PM}^S_{t-1}$ and can therefore execute independently. Their outputs converge at the retrieval stage: Intent determines \textit{what to search for} via the dual channels (Lines~21--23), while Reflection determines \textit{what to suppress} via the rank-cap (Line~24). TW-RRF (Line~26) then fuses the full trajectory stored in $\mathit{RM}$ into the final ranking, after which both memory components are updated to prepare the state for the next turn.

\section{Prompt Templates}
\label{sec:appendix_prompts}
To improve reproducibility, we provide the prompt templates used by the main reasoning modules of ReCoVR, together with the prompt template used by the simulated user. Tables~\ref{tab:prompt_intent}--\ref{tab:prompt_question} correspond to the four reasoning modules of ReCoVR (Intent Decomposer, Query Reformer, State Reflector, and Question Asking, respectively), following the execution order described in Algorithm~\ref{alg:recovr}. Table~\ref{tab:prompt_simulator} lists the prompts used by the simulated user (Sec.~\ref{sec:supp_sim_design}); all methods in our experiments share this same simulator configuration.

Two design principles run through all templates. First, each prompt enforces a strict \textit{role boundary}: modules are instructed to decompose, compress, or analyze without hallucinating details absent from the input, preventing error amplification across turns. Second, the Reflection prompt explicitly requests \textit{incremental} constraints, avoiding redundancy with the accumulated session state in $\mathit{PM}^S$.

\section{Limitations and Discussion}
\vspace{-1em}
\label{sec:appendix_limitation}

\minisection{Evaluation Scope and Interaction Variability.}
This work focuses on the algorithmic side of interactive retrieval: how feedback, retrieval history, and intermediate results should be used to update rankings across turns. Modeling the full diversity of real user behavior is a broader HCI-oriented research direction~\citep{balog2023user, kelly2009methods, marchionini2006exploratory,preece1994human}. It covers search strategies, patience, expertise, and interface effects. We therefore use a controlled interactive setting to isolate algorithmic differences under shared targets, initial rankings, feedback sources, and stopping criteria, and further complement it with a free-form human-feedback evaluation and degraded-feedback analysis. Larger-scale user studies with diverse behavioral profiles~\citep{balog2023user, zhang2024usimagent, mo2025survey} remain an important direction for future work. More broadly, ICoVR inherits the open-ended nature of composed retrieval, where interaction progressively reduces ambiguity among plausible targets rather than eliminating it entirely.

\minisection{Design Trade-offs and Deployment.}
ReCoVR uses simple, training-free, rank-level operations to remain backbone-agnostic and easy to plug into existing retrievers, and our ablations and sensitivity analyses suggest that the gains are not tied to narrow hyperparameter choices. Future work could make fusion more adaptive by estimating per-turn channel reliability from cross-channel agreement, query--reference consistency, or uncertainty signals. ReCoVR also introduces additional per-turn reasoning cost for intent decomposition and trajectory reflection; although this is partly offset by fewer executed turns, real-time deployment could benefit from caching, smaller VLMs, distillation, or lightweight structured parsers.

\minisection{Scope of the Reported Evidence.}
We summarize how each analysis should be interpreted.The paired intervals of Table~\ref{tab:paired_ci} quantify test-session sampling variability under a fixed model, simulator, and protocol; they are not tests over independent turns, and we report no $p$-values. The failure diagnostics of Sec.~\ref{sec:failure_diag} are temporal associations measured on executed trajectories: they show that ReCoVR recovers where the open-loop baseline stalls, but a single run cannot attribute that recovery to one module in isolation, which is why component attribution rests on the matched ablations instead. The rank-worsening row is an indicator based on target rank, not a full semantic measurement of query drift. The structured-output statistics describe interface reliability rather than the semantic correctness of each decision. Natural extensions include reasoning models below the 4B scale, user studies with broader participant pools, and latency measurements with question generation disabled. We do not report an oracle anchor, since it would require target-side information that is unavailable at retrieval time.

\minisection{Relation to LLM Agent Frameworks.}
ReCoVR shares structural similarities with agent frameworks such as ReAct~\citep{yao2022react} and Reflexion~\citep{shinn2023reflexion}, but is purpose-built for interactive retrieval. Its action space is constrained to query reformulation, dual-channel retrieval, rank-level intervention, and rank fusion, while its reflection is grounded in ranked-list trajectories rather than free-form critique. Reasoning traces may help single-turn edit understanding, but do not directly specify how to monitor drift, suppress repeated failures, or fuse evolving evidence across turns.

\minisection{Broader Impacts.}
This work may improve interactive access to large-scale video collections in applications such as media search, education, and assistive information retrieval. However, stronger video retrieval systems should be deployed carefully when applied to sensitive or personal video collections, where unauthorized search or unintended exposure may raise privacy concerns. Responsible deployment should consider data consent, access control, and dataset/model biases.

\begin{table*}[t]
\caption{\textbf{Prompt template for the Intent Decomposer} (Sec.~\ref{sec:dual_pathway} of the main paper). Given user feedback $u_t$, the module decomposes it into a target-descriptive search query and an optional edit instruction, corresponding to the T2V and CoVR channels respectively.}
\label{tab:prompt_intent}
\centering
\small
\setlength{\tabcolsep}{6pt}
\renewcommand{\arraystretch}{1.15}
\begin{tabular}{p{0.17\linewidth}p{0.77\linewidth}}
\toprule
\textbf{Field} & \textbf{Content} \\
\midrule
System prompt & \texttt{null} \\
Template &
\texttt{Task: Decompose user feedback into search and edit components for interactive video retrieval.}\\
&
\texttt{Context:}\\
&
\texttt{- Previous question asked to user: \{prev\_question\}}\\
&
\texttt{- Previous top-1 retrieved video description: \{prev\_top1\_caption\}}\\
&
\texttt{- User feedback: \{user\_query\}}\\
&
\texttt{Role boundary:}\\
&
\texttt{- Only decompose intent.}\\
&
\texttt{- Do NOT rewrite, enrich, or add details not present in feedback.}\\
&
\texttt{Steps:}\\
&
\texttt{1. Determine whether the feedback mainly answers the previous question.}\\
&
\texttt{2. Determine whether the feedback contains explicit edit intent}\\
&
\texttt{(change/remove/replace/add/make/swap/turn/switch/modify, etc.).}\\
&
\texttt{3. Extract search\_query\_info as target-descriptive information}\\
&
\texttt{(objects/actions/scenes) when available.}\\
&
\texttt{4. Extract edit\_query\_info only when edit intent exists.}\\
&
\texttt{5. If feedback is purely an answer and no clear standalone target}\\
&
\texttt{description is provided, search\_query\_info can be empty.}\\
&
\texttt{6. If no edit intent, set edit\_query\_info to an empty string.}\\
&
\texttt{Return JSON only: \{output\_schema\}} \\
Output schema &
\texttt{\{"reasoning": "string (1-2 sentences)",}\\
&
\texttt{"is\_answer\_to\_prev\_question": "boolean",}\\
&
\texttt{"has\_edit\_intent": "boolean",}\\
&
\texttt{"search\_query\_info": "string", "edit\_query\_info": "string"\}} \\
\bottomrule
\end{tabular}
\end{table*}

\begin{table*}[t]
\caption{\textbf{Prompt templates for the Query Reformer} (Sec.~\ref{sec:dual_pathway} of the main paper). Two sub-modules serve different purposes: \texttt{t2v\_initialize} generates the initial T2V search query at Turn~0 by applying $u_0$ to the reference video, while \texttt{rewrite\_queries} compresses the accumulated running query when it exceeds the length threshold at Turn~$t \geq 1$.}
\label{tab:prompt_reform}
\centering
\small
\setlength{\tabcolsep}{6pt}
\renewcommand{\arraystretch}{1.15}
\begin{tabular}{p{0.17\linewidth}p{0.77\linewidth}}
\toprule
\textbf{Field} & \textbf{Content} \\
\midrule
Module & \texttt{t2v\_initialize} (Turn~0 only) \\
System prompt & \texttt{null} \\
Template &
\texttt{Task: Generate a text-to-video search query from a reference video and an edit instruction.}\\
&
\texttt{You see a reference video. The user wants the target video after applying}\\
&
\texttt{the edit instruction.}\\
&
\texttt{Edit instruction: \{edit\_query\}}\\
&
\texttt{Requirements:}\\
&
\texttt{1. Use only visible video content plus the given edit instruction.}\\
&
\texttt{2. Produce one concise target description for retrieval.}\\
&
\texttt{3. Keep key visual details (objects, actions, scene, attributes)}\\
&
\texttt{while avoiding unnecessary verbosity.}\\
&
\texttt{Return JSON only: \{output\_schema\}} \\
Output schema &
\texttt{\{"reasoning": "string (1 sentence)", "search\_query": "string"\}} \\
\midrule
Module & \texttt{rewrite\_queries} (Turn~$t \geq 1$, triggered when query exceeds length threshold) \\
System prompt & \texttt{null} \\
Template &
\texttt{Task: Compress a long retrieval query while preserving all critical visual information.}\\
&
\texttt{Query to compress: \{query\_to\_rewrite\}}\\
&
\texttt{Requirements:}\\
&
\texttt{1. Preserve key objects, actions, scenes, attributes, and constraints.}\\
&
\texttt{2. Remove redundancy and keep retrieval semantics intact.}\\
&
\texttt{3. Return one concise rewritten query.}\\
&
\texttt{Return JSON only: \{output\_schema\}} \\
Output schema &
\texttt{\{"reasoning": "string (1 sentence)", "rewritten\_query": "string"\}} \\
\bottomrule
\end{tabular}
\end{table*}

\begin{table*}[t]
\caption{\textbf{Prompt template for the State Reflector} (Sec.~\ref{sec:dual_pathway} of the main paper). The \texttt{reflection\_context} variable is populated from $\mathit{PM}^S_{t-1}$ (accumulated queries, prior positive/negative constraints) and relevant gallery captions from $\mathit{PM}^L$, corresponding to the inputs $(u_t, \mathit{PM}^S_{t-1}, \mathit{PM}^L)$ in Eq.~(\ref{eq:reflect}).}
\label{tab:prompt_reflect}
\centering
\small
\setlength{\tabcolsep}{6pt}
\renewcommand{\arraystretch}{1.15}
\begin{tabular}{p{0.17\linewidth}p{0.77\linewidth}}
\toprule
\textbf{Field} & \textbf{Content} \\
\midrule
System prompt & \texttt{null} \\
Template &
\texttt{Task: Analyze user feedback on retrieval results to assess satisfaction}\\
&
\texttt{and extract incremental constraints.}\\
&
\texttt{Input feedback: \{user\_feedback\}}\\
&
\texttt{Context: \{reflection\_context\}}\\
&
\texttt{Instructions:}\\
&
\texttt{1. Decide satisfaction\_level in \{negative, neutral\}.}\\
&
\texttt{2. Use "negative" only for explicit rejection (e.g., wrong/not this/not desired).}\\
&
\texttt{3. Extract NEW positive constraints (what user wants) as short phrases.}\\
&
\texttt{4. Extract NEW negative constraints (what user wants to avoid) as short phrases.}\\
&
\texttt{5. Keep outputs incremental: do not repeat constraints already implied by context}\\
&
\texttt{unless the user explicitly strengthens them.}\\
&
\texttt{6. Do not hallucinate unstated details.}\\
&
\texttt{Return JSON only: \{output\_schema\}} \\
Output schema &
\texttt{\{"reasoning": "string (1-2 sentences)",}\\
&
\texttt{"satisfaction\_level": "negative | neutral",}\\
&
\texttt{"positive\_constraints": ["string"], "negative\_constraints": ["string"]\}} \\
\bottomrule
\end{tabular}
\end{table*}

\begin{table*}[t]
\caption{\textbf{Prompt template for Question Asking} (Sec.~\ref{sec:dual_pathway} of the main paper). The module identifies remaining ambiguities from the current session state and generates one clarifying question per turn.}
\label{tab:prompt_question}
\centering
\small
\setlength{\tabcolsep}{6pt}
\renewcommand{\arraystretch}{1.15}
\begin{tabular}{p{0.17\linewidth}p{0.77\linewidth}}
\toprule
\textbf{Field} & \textbf{Content} \\
\midrule
System prompt & \texttt{You generate one concise clarifying question for interactive video retrieval.} \\
Template &
\texttt{Current search query: "\{current\_search\_query\}"}\\
&
\texttt{Current edit query: "\{current\_edit\_query\}"}\\
&
\texttt{Current constraints:}\\
&
\texttt{- positive: \{constraints\_positive\}}\\
&
\texttt{- negative: \{constraints\_negative\}}\\
&
\texttt{Ask one short open-ended question that helps disambiguate remaining}\\
&
\texttt{uncertainty in target appearance, actions, or scene.}\\
&
\texttt{Return ONLY the question.} \\
Output schema &
\texttt{\{"reasoning\_brief": "string (1 sentence: what ambiguity you identified)",}\\
&
\texttt{"question": "string (one short question)"\}} \\
\bottomrule
\end{tabular}
\end{table*}

\clearpage
\begin{table*}[!ht]
\caption{\textbf{Prompt templates for the simulated user} (Sec.~\ref{sec:supp_sim_design}). The simulator follows a two-stage protocol: Stage~1 classifies the discrepancy between the target and the candidate (\texttt{MINOR\_DIFF} and \texttt{MAJOR\_DIFF} correspond to the \texttt{MINOR} and \texttt{MAJOR} status in Algorithm~\ref{alg:user_sim}, respectively); Stage~2 selects one of three feedback policies accordingly. All methods in our experiments share this identical simulator.}
\label{tab:prompt_simulator}
\centering
\small
\setlength{\tabcolsep}{6pt}
\renewcommand{\arraystretch}{1.15}
\begin{tabular}{p{0.17\linewidth}p{0.77\linewidth}}
\toprule
\textbf{Field} & \textbf{Content} \\
\midrule
System persona &
\texttt{You are a casual user searching for a video. You give short, direct feedback.}\\
&
\texttt{Do not explain your reasoning. Keep responses under 10 words.} \\
\midrule
Target caption prompt &
\texttt{Describe this target video in one short sentence. This is the video you should remember.} \\
\midrule
Stage 1: analysis &
\texttt{Compare two DIFFERENT videos step by step.}\\
&
\texttt{Video 1 = TARGET (what I want)}\\
&
\texttt{Video 2 = CANDIDATE (what was found)}\\
&
\texttt{Step 1: What is the KEY DIFFERENCE between the two videos?}\\
&
\texttt{Step 2: Do they share the SAME CATEGORY? (same animal type, same object type, same scene type)}\\
&
\texttt{Step 3: Decide status based on Step 2:}\\
&
\texttt{- If same category (e.g. goat vs goat, nuts vs nuts, river vs river) $\rightarrow$ MINOR\_DIFF}\\
&
\texttt{- If different categories (e.g. dog vs car) $\rightarrow$ MAJOR\_DIFF}\\
&
\texttt{Output JSON with fields in this order:}\\
&
\texttt{\{"diff\_analysis": "...", "same\_category": true/false,}\\
&
\texttt{"status": "MINOR\_DIFF" or "MAJOR\_DIFF"\}} \\
\midrule
Stage 2: modify &
\texttt{Video 2 is close but has specific errors (e.g., wrong color, wrong object).}\\
&
\texttt{Give a short command to fix it. Example: "Change the red car to a blue one"}\\
&
\texttt{or "Make it sunny".} \\
\midrule
Stage 2: rewrite &
\texttt{The candidate video is completely unrelated to the target. Ignore the candidate.}\\
&
\texttt{Write a new, stand-alone search query that describes Video 1 (Target).}\\
&
\texttt{Be descriptive but natural. Example: "A dog running on the beach under the sun".} \\
\midrule
Stage 2: answer &
\texttt{The search system asked: "\{question\}". Answer based on Video 1 (Target).}\\
&
\texttt{Answer directly and briefly. Example: "It is red", "He is running".}\\
&
\texttt{Do not speak in full paragraphs.} \\
\bottomrule
\end{tabular}
\end{table*}

\end{document}